\documentclass[5p,times,twocolumn,nopreprintline]{elsarticle}

\usepackage[utf8]{inputenc}
\usepackage{times}
\usepackage[english]{babel}
\usepackage{amsmath}
\usepackage{amssymb}

\usepackage{enumitem}
\usepackage{graphicx}
\usepackage{subfigure}
\usepackage{tabularx}
\usepackage{xspace}
\usepackage{xurl}
\setitemize{noitemsep,topsep=0pt,parsep=0pt,partopsep=0pt}
\usepackage{xcolor}
\usepackage{algpseudocode}
\usepackage{algorithm}
\usepackage[normalem]{ulem}

\usepackage[pagebackref=true,hyperfootnotes=false,colorlinks=true,linkcolor=red,urlcolor=blue,citecolor=blue]{hyperref}

\usepackage{shortcuts}

\usepackage[absolute,showboxes]{textpos}

\renewcommand{\paragraph}{\vspace{3pt}\noindent\textbf}
\begin{document}

\setlength{\TPHorizModule}{\paperwidth}
\setlength{\TPVertModule}{\paperheight}
\TPMargin{5pt}
\begin{textblock}{0.8}(0.1,0.02)
     \noindent
     \small
     \textcolor{blue!80!black}{If you cite this paper, please use the FGCS reference:
     Juan Caballero, Gibran Gomez, Srdjan Matic, Gustavo S{\'a}nchez, Silvia Sebasti{\'a}n, and Arturo Villaca{\~n}as.
     The Rise of GoodFATR: A Novel Accuracy Comparison Methodology for Indicator Extraction Tools.
     In \textit{Future Generation Computer Systems, Volume 144, 2023, Pages 74-89}.
     DOI: \url{https://doi.org/10.1016/j.future.2023.02.012}.}
\end{textblock}

\title{The Rise of GoodFATR: A Novel Accuracy Comparison Methodology\\
for Indicator Extraction Tools}

\author[1]{Juan Caballero}
\ead{juan.caballero@imdea.org}
\author[1,2]{Gibran Gomez}
\ead{gibran.gomez@imdea.org}
\author[1]{Srdjan Matic}
\ead{srdjan.matic@imdea.org}
\author[3]{Gustavo Sánchez}
\ead{sanchez@kit.edu}
\author[1,2]{Silvia Sebastián}
\ead{silvia.sebastian@imdea.org}
\author[1,2]{Arturo Villacañas}
\ead{arturo.villacanas@imdea.org}

\affiliation[1]{	organization={IMDEA Software Institute},
	addressline={Campus de Montegancedo, s/n},
	postcode={28223},
	city={Madrid},
	country={Spain}}

\affiliation[2]{	organization={Universidad Politecnica de Madrid},
	addressline={Campus de Montegancedo, s/n},
	postcode={28223},
	city={Madrid},
	country={Spain}}

\affiliation[3]{	organization={Karlsruhe Institute of Technology},
	addressline={Am Fasanengarten 5},
	postcode={76131},
	city={Karlsruhe},
	country={Germany}}

\begin{abstract}

To adapt to a constantly evolving landscape of cyber threats, organizations actively need to collect Indicators of Compromise (IOCs), 
i.e., forensic artifacts that signal that a host or network might 
have been compromised.
IOCs can be collected through open-source and commercial structured IOC feeds. 
But, they can also be extracted from a myriad of unstructured threat reports 
written in natural language and 
distributed using a wide array of sources such as blogs and social media.
There exist multiple indicator extraction tools that can 
identify IOCs in natural language reports. 
But, it is hard to compare their accuracy 
due to the difficulty of building large ground truth datasets.
This work presents a novel majority vote methodology 
for comparing the accuracy of indicator extraction tools, 
which does not require a manually-built ground truth.
We implement our methodology into \tool, an automated platform 
for collecting threat reports from a wealth of sources,
extracting IOCs from the collected reports using multiple tools, and 
comparing their accuracy.

\Tool supports 6 threat report sources:
\rss, \twitter, \telegram, 
\malpedia, \aptnotes, and \chainsmith.
\Tool continuously monitors the sources, 
downloads new threat reports, 
extracts \numiocs indicator types from the collected reports, and 
filters non-malicious indicators to output the IOCs.
We run \tool over 15 months to collect \numentries reports from the 6
sources; extract \numextractedindicators indicators from the reports; and
identify \numextractediocs IOCs.  We analyze the collected data to identify the
top IOC contributors and the IOC class distribution.
We apply \tool to compare the IOC extraction accuracy of 7 popular 
open-source tools with {\tool}'s own indicator extraction module.

\end{abstract}

\begin{keyword}
  Indicators of Compromise \sep IOC \sep Cyber Threat Intelligence \sep RSS \sep Twitter \sep Telegram
\end{keyword}

\maketitle

\section{Introduction}
\label{sec:introduction}

Cyber Threat Intelligence (CTI) provides information on attacker behavior 
that allows 
gaining visibility into the fast-evolving threat landscape; 
to understand the techniques, tactics, and procedures (TTPs) attackers use; and
to timely identify and contain attacks.
CTI is a multi-billion dollar industry, 
expected to keep growing to more than 16 billion USD by 2026~\cite{timarket}.
An essential piece of CTI is extracting and sharing 
\textit{indicators of compromise} (IOCs), 
forensic artifacts that when observed on a device or network indicate
it may have been compromised, 
e.g., malicious IPs, domains, and file hashes.
IOCs are an actionable piece of CTI, as they can be fed to security systems 
(e.g., NIDS, firewall, HIDS, blocklists) to detect and block attacks.

IOCs can be distributed through open and commercial 
feeds~\cite{tealeaves,bouwman2020different},
which provide structured IOC data 
following standardized formats 
(e.g., STIX~\cite{stix}, OpenIOC~\cite{openioc}).
However, much CTI is distributed through unstructured threat reports,
written in natural language and published through a wealth of 
security blogs 
and social media platforms.
For example, Twitter has become widely used 
for exchanging and spreading cybersecurity information, 
not only by cybersecurity companies 
but also by experts that sometimes rush to share 
their discoveries~\cite{sabottke_usenix2015,alves_esorics2020}.
Natural language reports often contain IOCs and 
typically provide more detailed contextual descriptions about the IOCs
compared to IOC feeds.
In addition, while IOC feeds usually focus on a small set of indicator types 
(i.e., IP addresses, domain names, file hashes), 
unstructured threat reports frequently provide more varied 
indicator types such as vulnerability identifiers, 
email addresses, blockchain addresses, Tor onion addresses,
fuzzy hashes (e.g., SSDeep~\cite{ssdeep}), social handles,
and target countries.

There exist multiple indicator extraction tools
that can identify IOCs from unstructured threat reports in natural 
language using \regexps~\cite{jager,iocparser,cacador,cyobstract,iocextract,iocextractor}.
Even recent works that apply natural language processing (NLP) techniques to 
analyze threat reports still use \regexps for identifying 
indicators as part of their pipeline~\cite{liao2016acing,ttpdrill,chainsmith,satvat2021extractor}.
\Regexps for extracting indicators can easily become complex, 
making it difficult to understand what they match. 
Also, small differences between \regexps for the same indicator type 
may significantly affect the extraction results.
Furthermore, \regexps can be affected by catastrophic backtracking 
introducing ReDoS vulnerabilities~\cite{davis2018impact}.
Despite the popularity of \regexp based indicator extraction 
tools~\cite{jager,iocparser,cacador,cyobstract,iocextract,iocextractor} 
no previous work has systematically evaluated them to understand 
which tool is more accurate for each indicator type and how accurate their 
extraction is. 
The main challenge to evaluate indicator extraction accuracy 
is the difficulty of building large-scale ground truth datasets 
from real threat reports.
To address this challenge, 
this work presents a novel majority vote methodology
for comparing the accuracy of indicator extraction tools,
which does not require a manually-built ground truth.

Another challenge is that in a large and fast-evolving landscape of threats, 
coverage is difficult to achieve by any single entity, 
as shown by different sources having little 
IOC overlap~\cite{tealeaves,bouwman2020different}.
Thus, it is not sufficient to rely on a single source, or a small set of 
sources, to build an accurate, comprehensive, and up-to-date IOC list.
To overcome this limitation, organizations and security analysts  
can resort to extracting IOCs from the threat reports collected from 
multiple sources.
However, there exists a myriad of sources through which threat reports 
are disseminated including 
hundreds of blogs, Twitter accounts, and Telegram channels.
To address this challenge, we present a modular threat report 
collection pipeline that can collect reports 
from a wealth of \emph{sources} in a unified manner.

Our collection pipeline currently supports 6 diverse sources:
\rss, \twitter, \telegram, and three report datasets 
\malpedia~\cite{malpedia}, \aptnotes~\cite{aptnotes}, and 
\chainsmith~\cite{chainsmith}.
For \rss, \twitter, and \telegram it takes as input a list of 
\emph{origins} (\rss feeds, \twitter accounts, \telegram channels) to monitor. 
It periodically visits those origins to identify new \emph{entries} 
(blog posts, tweets, messages). 
For entries that contain a URL to a report, it downloads the 
report's \emph{document}.
The selected sources are complementary. 
\rss allows collecting reports from hundreds of blogs from 
cybersecurity companies and individual experts, while \twitter and \telegram 
cover social media distribution. 
In addition, crowd-sourced datasets such as \malpedia and \aptnotes 
allow identifying previously unknown blogs and \twitter accounts 
that should be monitored, as well as 
collecting reports from blogs without an \rss feed.

We have implemented our novel methodology for comparing indicator 
extraction tools and our collection pipeline into an 
automated platform called \tool.
\Tool supports a variety of open-source indicator extraction tools 
but also provides its own \searcher tool, 
which takes as input a document in HTML, PDF, or plain text format and  
applies \regexps to extract \numiocs indicator types.
When processing reports, indicator extraction tools may identify 
malicious indicators (e.g., C\&C domains), as well as benign ones
(e.g., URL references or contact emails for security companies). 
We call malicious indicators IOCs and benign indicators \emph{generic}.
\Tool provides a filtering module that removes generic indicators 
extracted from the documents to output only the IOCs.
Its filtering module builds a dynamic blocklist from the monitored 
sources and origins.
This approach avoids hard-coded blocklists used by prior tools, 
which cannot adapt to the different sources and origins each user may monitor.

We use \tool to compare the accuracy of  
7 popular open-source indicator extraction tools~\cite{jager,iocparser,cacador,cyobstract,iocfinder,iocextract,iocextractor}, 
as well as {\tool}'s own \searcher tool, on \numcomparisonreports reports.
The results show \searcher being the most accurate tool on 11 of the 13
indicator types supported by multiple tools.
We also identify ReDos vulnerabilities in two of the tools, 
one new and one previously reported.
We also evaluate the filtering module provided by \tool 
against the filtering rules used in two prior tools.
The filtering by \tool achieves an F1 score of 0.91 compared to
0.47 (\iocparser) and 0.46 (\cacador).

We have used \tool to
collect \numentries reports over 15 months from \numorigins origins 
distributed through 6 sources;
extract \numextractedindicators indicators from the reports; and
identify \numextractediocs IOCs.
We analyze the collected data to identify the top IOC contributors and
the IOC class distribution.

This work provides the following contributions:

\begin{itemize}

\item We present a novel majority-vote methodology for evaluating the 
accuracy of indicator extraction tools, 
which does not require a manually-built ground truth. 
We apply our methodology to compare the accuracy of 7 popular tools and 
our own \searcher over \numcomparisonreports reports.
The results show \searcher is the most accurate tool on 11 of the 13 
indicator types supported by multiple tools. 

\item We present \tool, an automated platform to collect threat reports 
from a variety of sources in a unified manner, 
extract the indicators in the collected reports, and
filter generic indicators to output only the IOCs.

\item We use \tool to 
collect \numentries reports over 15 months from the 6 sources; 
extract \numextractedindicators indicators from the reports; and 
identify \numextractediocs IOCs.

\item We release the source code of the \searcher tool
at \url{https://github.com/malicialab/iocsearcher}

\end{itemize}

The remainder of this paper is organized as follows.
Section~\ref{sec:related} presents prior related work. 
Section~\ref{sec:overview} provides an overview of our platform.
Section~\ref{sec:collection} details the sources and approach used in the threat report collection.
Section~\ref{sec:tools} performs a systematization and comparative study of 8 indicator extraction tools.
Section~\ref{sec:evaluation} analyzes the reports collected over 15 months, the IOCs extracted from those reports, and 
the top contributing sources and origins.
Section~\ref{sec:evalTools} presents our novel accuracy comparison methodology and applies it to compare the 8 indicator extraction tools on \numcomparisonreports threat reports.
Section~\ref{sec:discussion} discusses limitations and future improvements.
Finally, Section~\ref{sec:conclusions} concludes.

\begin{table}[t]
\caption{\updated{Summary of prior academic works on threat report collection and IOC extraction. Open-source IOC extraction tools are summarized in Table~\ref{tbl:tools}}.}
\label{tbl:related}
\centering
\small
\begin{tabular}{l|c|c|c|c|c|c|c|}
\cline{3-8}
\multicolumn{2}{c|}{} & 
\multicolumn{4}{c|}{\bf Collection Sources} & 
\multicolumn{2}{c|}{\bf Extraction} \\ 
\hline
{\bf Work} & 
{\bf Year} & 
\featuretexta{Blogs - Crawlers } & 
\featuretexta{Blogs - RSS} & 
\featuretexta{Telegram} & 
\featuretexta{Twitter} & 
\featuretexta{Regexp} &  
\featuretexta{NLP} \\  
\hline
Liao et al.~\cite{liao2016acing} & 2016 & \Y & \N & \N & \N & \Y & \Y  \\ TTPDrill~\cite{ttpdrill} & 2017 & \N & \N & \N & \N & \Y & \Y \\ Malpedia~\cite{malpedia} & 2017 & \Y & \N & \N & \N & \Y & \N \\ \chainsmith~\cite{chainsmith} & 2018 & \Y & \N & \N & \N & \Y & \Y \\ IOCMiner~\cite{iocminer} & 2019 & \N & \N & \N & \Y & \Y & \N \\ TIMiner~\cite{timiner} & 2020 & \Y & \N & \N & \N & \Y & \Y \\ Alves et al.~\cite{alves_esorics2020} & 2020 & \N & \N & \N & \Y & \Y & \N \\ Twiti~\cite{twiti} & 2021 & \N & \N & \N & \Y & \Y & \Y \\ \hline
\tool & 2022 & \N & \Y & \Y & \Y & \Y & \N \\
\hline
\end{tabular}
\end{table}

\ignore{
\begin{table}[t]
  \caption{\updated{Comparison of previus work on IOC extraction tools and IOC data feeds}.}
  \label{tab:relatedWork}
  \resizebox{\columnwidth}{!}{    \begin{tabular}{c|c|c|c|c|c|c|}
      \cline{2-6}

      \multicolumn{1}{c|}{System} & \begin{tabular}[c]{@{}c@{}}Release \\ Date\end{tabular} & \begin{tabular}[c]{@{}c@{}}Extraction \\ Method\end{tabular} & \begin{tabular}[c]{@{}c@{}}Crawler \\ Sources\end{tabular} & \# Sources & \begin{tabular}[c]{@{}c@{}}Tool \\ Available\end{tabular}\\
      \hline
      \jager~\cite{jager}                   & 2015 & regex & \N & 0 & \Y\\
      \iocparser~\cite{iocparser}           & 2017 & regex & \N & 0 & \Y\\
      \cacador~\cite{cacador}               & 2016 & regex & \N & 0 & \Y\\
      \cyobstract~\cite{cyobstract}         & 2018 & regex & \N & 0 & \Y\\
      \iocextract~\cite{iocextract}         & 2019 & regex & \N & 0 & \Y\\
      \iocextractor~\cite{iocextractor}     & 2019 & regex & \N & 0 & \Y\\
            Liao et al.~\cite{liao2016acing}      & 2016 & NLP + regex & Techical blogs & 45 & \N\\
      \chainsmith~\cite{chainsmith}         & 2018 & regex & Security Articles (i.e., blogs) & 10 & \N \\
      TTPDrill~\cite{ttpdrill}              & 2017 & NLP & \N & 0 & \N\\
      TIMiner~\cite{timiner}                & 2020 & CNN & Security blogs, Security vendor bulletins, hacking forums & 75 & \Y\\
            Alves et al.~\cite{alves_esorics2020} & 2020 & - & - & 0 & \N\\
      \#Twiti~\cite{twiti}                  & 2021 & & Twitter API, Twitter accounts, Security blogs & ?, 146, 10 & \Y\\
            IOCMiner~\cite{iocminer}             & 2019 & regex & Twitter accounts& 75 & \Y \\
      Guo Li et al~.\cite{tealeaves}       & 2019 & \N & Facebook ThreatExchange, Paid Feed Aggregator, Paid IP Reputation System, Public Blocklist & ??? & \N\\ 
      \searcher & 2022 & regex & APTnotes, ChainSmith, Malpedia, RSS, Twitter accounts, Telegram & 3,226 & \Y\\
      \hline

    \end{tabular}  }
\end{table}
}

\section{Related Work}
\label{sec:related}

\updated{
Extracting IOCs from natural language documents is an important CTI task
that comprises two problems: threat report collection and IOC extraction.
Prior work can be split into two groups.
First, there exist open source tools that focus on extracting IOCs 
from a given threat report using \regexps 
(\eg~\cite{jager,iocparser,cacador,cyobstract,iocextract,iocextractor}). 
These tools do not address the problem of threat report collection. 
We summarize the most popular open-source IOC extraction tools 
in Table~\ref{tbl:tools} and detail how they work in Section~\ref{sec:tools}.
Second, prior academic works typically address both 
the threat report collection and the IOC extraction problems.
We detail prior academic work in the remainder of this section.

\paragraph{Threat report collection.}
Table~\ref{tbl:related} shows that prior academic work has collected 
reports from one source, either from 
cybersecurity blogs~\cite{chainsmith,liao2016acing,timiner,malpedia} or 
from Twitter~\cite{iocminer,alves_esorics2020,twiti}.
There is also one work that does not address threat report collection, 
focusing exclusively on IOC extraction~\cite{ttpdrill}.
In contrast, our work proposes a modular threat report collection platform 
that can collect reports from multiple sources, 
namely cybersecurity blogs, \telegram, \twitter, and 
datasets provided by prior works 
such as \chainsmith~\cite{chainsmith}, \malpedia~\cite{malpedia}, and 
\aptnotes~\cite{aptnotes}.

Previous work that collected threat reports from cybersecurity blogs 
built dedicated crawlers for each blog of interest
(e.g.,~\cite{chainsmith,liao2016acing,timiner}).
The advantage of dedicated crawlers is that they can be designed to collect the
historical archive of reports available on a blog's website.
One disadvantage is that a crawler needs to be developed for each blog, which
limits the number of monitored blogs.
For example, Liao et al.~\cite{liao2016acing} collected reports from 45 blogs,
Zhu and Dumitras from 10~\cite{chainsmith}, and Zhao et al.~\cite{timiner} from 75.
Moreover, as websites evolve, the crawlers may break and need to be updated.
For example, \chainsmith originally collected reports from 10 blogs in 2018, but the
supported blogs gradually decreased over time, possibly due to
crawlers breaking due to website updates.
In mid-2020 only two blogs were still being crawled, and since mid-2021
\chainsmith does not collect any new blog posts.

Rather than building dedicated crawlers, 
\tool collects threat reports from cybersecurity blogs that offer an RSS feed.
The use of RSS feeds allows us to build a generic collection module for 
cybersecurity blogs, enabling us to scale the collection to  
\rssOrigins blogs, 6 times larger than prior works.
Furthermore, our RSS module still collects \entries from the same set of 
blogs supported by \chainsmith, showing that those \origins are still active.
It is important to note that our RSS feed collection may also require updates
(\eg if a feed's XML URL changes), 
but updating a feed's URL is significantly easier than 
building a new dedicated crawler.
Furthermore, if a blog does not offer an RSS feed, 
it will still need a dedicated crawler.

\paragraph{IOC extraction.}
All open-source IOC extraction tools in Table~\ref{tbl:tools} and 
three academic works 
in Table~\ref{tbl:related}~\cite{malpedia,iocminer,alves_esorics2020}
extract IOCs from the collected reports using \regexps.
Some academic papers have instead proposed 
natural language processing (NLP) approaches to improve 
the extraction~\cite{liao2016acing,chainsmith,ttpdrill,timiner}.
However, NLP approaches also rely on \regexps to extract indicators in 
their pipelines.
Despite the popularity of \regexp-based IOC extraction, 
the large number of tools a user can choose for this task, and 
the impact of \regexp differences in the extraction results, 
we are not aware of any prior work that systematically analyzes 
which tool works for each indicator type.
Such a comparison should include the \regexp used for the extraction, 
but also other important steps such as 
how to filter benign indicators that are not IOCs.
The main difficulty for such evaluation is building large-scale datasets 
from real documents.
To address this challenge, in this work we propose a novel methodology to 
compare indicator extraction tools, which does not require a 
manually-built ground truth.

Also related are metrics to evaluate the quality of IOC feeds~\cite{tealeaves}.
We apply such metrics to compare different sources and origins to identify the
best and worst IOC contributors.
}

\paragraph{Extended version.}
This paper is an extended version of a previous 4-page work-in-progress
paper that appeared in JNIC 2022~\cite{caballero_jnic2022}.
Compared to that work, this paper provides four new contributions.
First, we double the number of sources from which we collect reports by
including three curated sources specialized in threat reports (\malpedia,
\aptnotes, \chainsmith).  We also expand the collection period to 15 months
(Section~\ref{sec:collection}).
Second, we provide an in-depth comparative analysis of 7 popular 
indicator extraction tools, as well as our own \searcher tool
(Section~\ref{sec:tools}).
Third, we redo our analysis of the collected data including the new sources and
the extended observation period, and add a new analysis of the distribution of
IOCs and the top IOC contributors (Section~\ref{sec:evaluation}).
Fourth, we develop a novel majority-vote methodology to automatically 
compare the accuracy of indicator extraction tools.
We apply this methodology to the 8 indicator extraction tools 
(Section~\ref{sec:evalTools}).

\begin{figure}[t]
  \centering
  \includegraphics[width=\columnwidth]{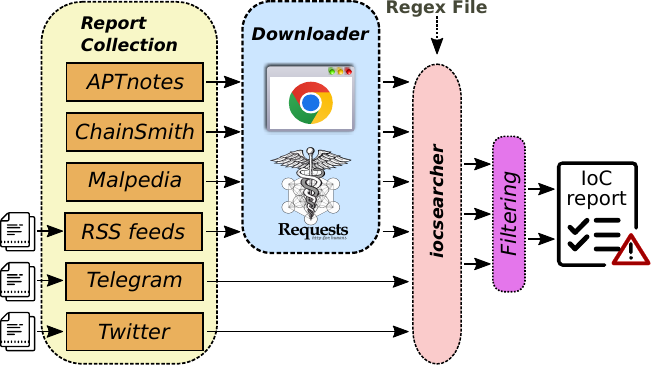}
  \caption{The architecture of \tool.}
  \label{fig:platform}
\end{figure}

\section{Overview}
\label{sec:overview}

An \emph{indicator} represents an artifact such as 
an IP address, a domain name, a URL, a file, a registry key, 
or a blockchain address.
An indicator is typically represented as a pair of a type (e.g., \ioc{email})
and a string value (e.g., \url{contact@test.com}).
An \emph{indicator of compromise} (IOC) is a malicious indicator whose 
presence on a device or network indicates the device might have been 
compromised.

This paper presents \tool, a platform for collecting threat reports 
from a wealth of sources and extracting IOCs from 
the collected reports.
Figure~\ref{fig:platform} depicts the architecture of \tool.
It consists of four modules: 
\textit{Report Collection}, \textit{Downloader}, \Searcher, and 
\textit{Filtering}.
The Report Collection gathers documents from a wealth of sources
including RSS feeds, Telegram channels, Twitter accounts, and 
threat report datasets
like Malpedia~\cite{malpedia}, APTnotes~\cite{aptnotes}, and
ChainSmith~\cite{chainsmith}.
The collection outputs a list of \emph{entries} each 
corresponding to the observation of a report in a source. 
Each entry contains the URL of the report 
or the content of a tweet or Telegram message. 
In addition, it may contain optional metadata about the report 
(e.g., publication date, author, language) and 
the source where it was observed 
(e.g., RSS feed, Twitter account).
The Downloader leverages an instrumented Chrome browser 
(and Python's requests library as backup)
to fetch the document pointed by an entry's URL and store it to file 
together with download metadata
(e.g., download date, HTTP status code, redirection chain followed).
For Twitter and Telegram, the Downloader is not used, 
as their entries already contain the content of the posts.
\searcher takes as input a document in HTML, PDF, or plain text format, 
extracts its text (for HTML and PDF documents), and 
applies \regexps to identify \numiocs indicators.
We choose \regexps because they are an efficient technique for identifying, 
in a given string, indicators with some intrinsic structure 
such as email addresses or URLs.
To extract indicators, \searcher applies a match-and-validate approach. 
Matching applies each of the \regexps to identify candidate indicators 
in the input string.
Validation uses a function specific to each indicator type to ensure 
the candidate is indeed an indicator.
Separating matching from validation allows the validation to perform 
computations that are not possible within \regexps such as checking 
a checksum embedded in an indicator value 
(e.g., blockchain addresses, bank account numbers).
It also prevents \regexps from becoming too complex.
While validation minimizes the set of incorrect indicators output, 
it is possible that indicators present in the text are not IOCs, 
but rather correspond to benign indicators 
(e.g., the email of the report's author).
Thus, before generating the final list of IOCs, 
the Filtering removes generic indicators 
that are not IOCs.

Throughout its pipeline, \tool maintains traceability. 
From an IOC an analyst can identify the documents that contained it, 
from a document the entries from where it was collected, and 
from an entry the sources where it was observed.

\section{Threat Report Collection}
\label{sec:collection}

Our threat report collection module has been designed to support a variety of 
threat report sources.
We collect information from RSS, Twitter, Telegram, and publicly available 
report datasets such as the \malpedia malware encyclopedia~\cite{malpedia}, the \aptnotes repository~\cite{aptnotes}, and the \chainsmith database~\cite{chainsmith}.
\updated{
To handle diverse sources in a unified manner, 
the collection is structured around three concepts: 
\emph{\origin}, \emph{\entry}, and \emph{\doc}.

An \origin captures the distribution vector through which a report is 
disseminated, at a finer granularity than a source.
Each source typically has many \origins; 
for RSS, an \origin corresponds to the feed from a specific blog, 
for Twitter to a user account, and 
for Telegram to a channel. 
For report datasets (\malpedia, \aptnotes, \chainsmith) 
we use as \origin the organization that authored the report 
(e.g., Norton, TrendMicro). 
The same threat report might be distributed through multiple origins 
within the same source (e.g., through different blogs), 
as well as through different sources 
(e.g., a blog and a Twitter account).
We use the \origins to produce fine-grained statistics about the 
report distribution, 
e.g., to measure which Twitter accounts, 
Telegram channels, and 
RSS feeds provide more threat reports.
We format the origin to also include the source
in the form \emph{source:\origin}. 
For example, \emph{rss:nakedsecurity} and \emph{twitter:nakedsecurity} 
correspond to the RSS feed of the Naked Security blog by Sophos, and 
its Twitter account \url{@NakedSecurity}, respectively.

An \entry captures a specific mention of a threat report through an \origin. 
It corresponds to a post in a Twitter account, Telegram channel, or RSS feed.
For report datasets, an \entry is a record in the report metadata, namely a line in the report index file of \aptnotes, a row in the database of \chainsmith, and a BibTex entry in the \malpedia bibliography.
Each \entry has an \origin and contains either a download URL pointing to the report or in the case of social media, a string with the post's text.
An \entry may also contain additional report metadata as provided by the source:
the report author, its original publication timestamp, and 
the last update timestamp.

A \doc is an HTML, PDF, or text file that has been collected by \tool. 
A \doc is identified by the file's SHA256 hash. 
A \doc is generated every time the Downloader downloads a URL and also
from the text content of each \twitter and \telegram entry.
A \doc corresponds to an instance of a threat report.
The distinction between \emph{threat report} and \emph{document} is 
important to understand how \tool handles report distribution and 
changes to the report over time.
Multiple \docs collected by \tool may correspond to the same report.
There are three main reasons for this. 
First, a threat report may be distributed in different formats.
For example, an APT report originally distributed as a webpage in a 
blog will be stored by \aptnotes as a PDF document. 
When \tool collects the report from the blog's RSS feed 
it will obtain an HTML document.
When \tool collects the report from \aptnotes, it will obtain a PDF document.
Both documents are different instances of the same threat report.
Second, for reports distributed through URLs, 
if \tool downloads the URL at different points in time, 
it might collect a different HTML document (i.e., different SHA256) each time.
This may happen if the webpage contains dynamic content that changes over time. 
Each of those HTML documents is an instance of the same threat report.
Third, a threat report may have multiple versions. 
For example, the original blog entry for the report may be updated a day later 
to fix an errata. 
If \tool downloads the URL of the report on both days, it will collect two 
different HTML documents, which correspond to different versions of the 
same threat report. 

Identifying which collected documents 
correspond to the same threat report is not necessary for 
the operation of \tool.
All collected documents go through IOC extraction and 
the extracted IOCs are aggregated and deduplicated. 
If two documents are different versions of the same report, 
\tool will extract the same (or very similar) IOCs from those documents 
and will remove the duplicated IOCs. 
\Tool provides information that can help identify documents that 
correspond to the same report, 
e.g., by querying for all documents downloaded from the same URL or 
by searching for all documents with the same title 
(extracted from the PDF and HTML document metadata). 
But, this problem is not addressed in this paper.
We further discuss this issue in Section~\ref{sec:discussion}.

}

\subsection{Threat Report Sources}
\label{sec:sources}

\begin{table}[t]
  \small
  \centering
  \caption{Sources used for collecting threat reports.}
  \label{tbl:sources}
  \begin{tabular}{llccccc}
    \textbf{Source} & \textbf{Origin} & \featuretexta{Input Origins} & \featuretexta{Cumulative} & \featuretexta{Download URL} & \featuretexta{Content} & \featuretexta{IOCs} \\
    \hline
    aptnotes & organization & \N & \Y & \Y & \Y & \N \\
    chainsmith & organization & \N & \Y & \Y & \N & \Y \\
    malpedia & organization & \N & \Y & \Y & \N & \N \\
    rss & feed & \Y & \N & \Y & \N & \N \\
    telegram & channel & \Y & \Y & \N & \Y & \N \\
    twitter & account & \Y & \Y & \N & \Y & \N \\
    \hline
  \end{tabular}
\end{table}

Table~\ref{tbl:sources} summarizes the six report sources currently supported by \tool. 
Each source has a dedicated collection submodule. 
The RSS, Twitter, and Telegram submodules take as input a file that specifies the \origins (i.e., feeds, accounts, channels) that should be monitored. 
The report datasets do not require an input origin list as the whole dataset is downloaded.
We manually created our initial \origin lists for RSS, Twitter, and Telegram by including resources from prominent companies, cybersecurity news websites, 
and well-known security experts. 
We continuously add new \origins as we identify interesting 
RSS feeds, Twitter accounts, and Telegram channels.
We discuss the update process in Section~\ref{sec:discussion}.
Table~\ref{tbl:collection} captures the total number of \origins in each source at the time of writing:
\rssOrigins RSS feeds,
\twitterOrigins Twitter accounts,
\telegramOrigins Telegram channels,
\malpediaOrigins organizations in \malpedia, 
\aptnotesOrigins organizations in \aptnotes, and
\chainsmithOrigins blogs in \chainsmith.

For each source, Table~\ref{tbl:sources} shows how the \origin is defined and 5 properties: whether it requires an input list of \origins; whether the reports are cumulative (i.e., old \entries are always available) or \entries expire after some time; whether it provides a URL for each \entry from where the report needs to be downloaded; whether it includes the actual report or only the URL to the report; and whether it includes IOCs extracted from the report.

\paragraph{RSS.}
We use RSS feeds to collect reports from \rssOrigins security blogs.
Each RSS feed is uniquely identified by a URL pointing to the feed's XML file.
The owner of an RSS feed configures it to provide a maximum number of 
the latest entries, \eg the 5, 10, or 100 most recent ones.
To avoid missing entries, the RSS module needs to visit a feed 
frequently enough that the feed has not posted more entries than the 
maximum available since the RSS module last visited it.
For example, the \emph{Hexacorn} blog only provides the last 5 posts in 
its feed, but they only post twice per month. 
Thus, visiting its feed every two months would be enough to avoid 
missing entries. 
For each feed, the RSS module monitors the maximum number of entries 
it has obtained in a visit 
(\ie the configured maximum unless the feed has posted
less than the maximum since its creation), 
as well as the maximum daily rate at which posts are added.
While we could configure a different visit frequency for each feed, 
we have empirically observed that no feed posts more than their configured 
maximum in a single day. 
Thus, for simplicity, the RSS module queries each feed on a daily basis.
The module is configured to throw a warning if any feed publishes more than its 
maximum number of available entries in a single day. 
So far, the warning has not been observed, so no entries have been missed.

Since the RSS module is configured not to miss feed entries, 
two consecutive visits to the same feed will provide some overlapping entries.
The RSS module has an incremental approach where feed entries that 
were already collected (\ie older than the last feed visit) are ignored.
Only the new \entries are passed to the \downloader to collect their content.
The incremental approach avoids downloading the same content multiple times.

\paragraph{Telegram \& Twitter.}
The Telegram and Twitter modules leverage the official APIs to query each service.
Using these APIs they can fetch not just the most recent messages, but any message that was ever posted on an account or channel.
Each module takes as input a list of \origins 
that should be monitored, and connects to each \origin on a daily basis to fetch all new posts.
Using the API, our modules collect each post from an account, or channel, only once.
However, it is possible to obtain posts with the same content from multiple \origins, e.g., if different users re-tweet the same original message.

\paragraph{\malpedia.}
\malpedia offers a manually-curated BibTex file with reports related to malware.
BibTex \entries for new reports are added on a 
nearly daily basis~\cite{malpedia}. 
The BibTex file is generated dynamically upon request and includes all reports in the database in a cumulative manner.
Each bibliography \entry contains the following report fields: author, title, publication date, organization (which we use as \origin), URL, language, and the date when the report was added to \malpedia.
The reports' \docs are not provided, so the URLs are passed to the \downloader to collect them.

\paragraph{\aptnotes.}
\aptnotes is a GitHub repository with manually-curated APT reports~\cite{aptnotes}. 
It provides an index file with report metadata
containing the report's title, source (i.e., author organization which we use as \origin), publication date, the filename and file hash of the report \doc, and the URL where the \doc can be obtained.
Reports in webpages are converted from HTML to PDF.
All reports are stored in the \url{box.com} service as PDF files.

\paragraph{\chainsmith.}
The \chainsmith project~\cite{chainsmith} makes available an SQLite3 database with the reports it has downloaded, their metadata, and the IOCs it has extracted from the reports.
The database has been updated on a weekly basis since 2018.
It contains one table where each row corresponds to an IOC extracted from a report including the report URL, the source (i.e., the blog identifier we use as the \origin), the report title, and the publication date. 
The reports' \docs are not provided, so we extract the \doc URLs and pass them to \downloader.

\subsection{Downloader}
\label{sec:downloader}
The \downloader takes as input an \entry and tries to download the content pointed to by the \entry's URL.
It uses Selenium, a popular framework used for testing Web applications~\cite{selenium}.
We instrument Selenium to render URLs using a fully-fledged instance of Google Chrome.
The \downloader is able to follow redirects, it supports dynamic content executed with JavaScript, and, in addition to HTML pages, it can also download plain text documents as well as other MIME types such as PDFs.
In case our instrumented browser did not succeed in retrieving the content, the \downloader makes an additional attempt with Python's \textit{requests} library~\cite{python_requests}.
For each successfully downloaded URL, the \downloader stores the \doc to a file and it updates the \entry with the download information including the download timestamp, the \doc hash, the HTTP status code, and the redirection chain followed. 
In the final step, the downloaded \docs are filtered to remove resources with HTTP status code errors and HTML content where the title states the webpage was not found.

\begin{table*}[t]
\small
\centering
\caption{Comparison of indicator extraction tools.}
\label{tbl:tools}
\begin{tabular}{llr|ccc|cccr|c}
\cline{4-10}
\multicolumn{3}{c|}{} & \multicolumn{3}{c|}{\textbf{Inputs}} & \multicolumn{4}{c|}{\textbf{Indicator Extraction}} & \\
\hline
\textbf{Tool} & \textbf{Language} & \textbf{Stars} & \textbf{Text} & \textbf{HTML} & \textbf{PDF} & \textbf{Rearm} & \textbf{Validate} & \textbf{Dedup.} & \textbf{IOCs} & \textbf{Filtering}\\
\hline
\Jager~\cite{jager} & Python & 69 & \CC & \CC & \CC & \cc & \cc & \CC & 11 & \cc \\
\Iocparser~\cite{iocparser} & Python & 378 & \CC & \CC & \CC & \lc & \cc & \lc & 11 & \CC \\
\Cacador~\cite{cacador} & Go & 119 & \CC & \cc & \cc & \lc & \cc & \CC & 12 & \lc \\
\Cyobstract~\cite{cyobstract} & Python & 66 & \CC & \cc & \cc & \CC & \lc & \cc & 25 & \cc \\
\Iocfinder~\cite{iocfinder} & Python & 96 & \CC & \cc & \cc & \CC & \cc & \lc & 25 & \cc \\
\Iocextract~\cite{iocextract} & Python & 356 & \CC & \cc & \cc & \CC & \cc & \cc & 8 & \cc \\
\Iocextractor~\cite{iocextractor} & JavaScript & 35 & \CC & \cc & \cc & \CC & \cc & \CC & 18 & \cc \\
\hline
\Searcher~\cite{iocsearcher} & Python & - & \CC & \CC & \CC & \CC & \CC & \CC & \numiocs & \CC \\
\hline
\end{tabular}
\end{table*}

\section{Indicator Extraction: A Comparative Study}
\label{sec:tools}
In this section, we perform a survey and functionality comparison of indicator extraction tools including our own. 
We further quantitatively evaluate the tools in Section~\ref{sec:evalTools}.

To identify the tools, we search for open-source projects for extracting indicators. 
For this, we query GitHub for projects related to the \emph{ioc} keyword. 
Then, we manually examine each matching project to select those that correspond to tools that extract indicators. 
We keep only popular tools with at least 30 stars, which helps avoid the forks of more popular projects.
If an identified tool references any other indicator extraction tools, we also include those in our search.
This process identifies \numavailabletools popular open-source tools for extracting indicators. 
We also include our own \searcher tool, for a total of \numtools tools in Table~\ref{tbl:tools}.
 
The \numtools tools follow the same model comprising three steps, two of which are optional. 
The first optional step is \emph{text extraction}, which given an HTML or PDF document, extracts its text. 
The second step, present in all tools, is \emph{indicator extraction}, which extracts the indicators present in an input string by applying a number of \regexps, grammars, and rules.
Finally, some tools apply an optional \emph{filtering} step whose goal is to remove indicators that are benign (e.g., domains of security vendors), and thus cannot be considered IOCs.
The dominant language for implementing the tools is Python used by 6 tools including the two most popular tools (\iocparser and \iocextract) and our own \searcher tool. 
\Cacador is written in Go and \iocextractor in JavaScript.

In Table~\ref{tbl:tools}, a solid circle (\CC) indicates full support, and a half-filled circle (\lc) partial or optional support, and an empty circle (\cc) no support.
Next, we detail each of the three steps.

\subsection{Text Extraction}
\label{sec:textextract}
Since most threat reports are distributed as HTML and PDF documents, text extraction is an important step in the complete indicator extraction process.
However, it can be performed as a separate pre-processing step that takes as input an HTML or PDF document and outputs its content as text.
Content extraction is an independent process from identifying indicators, and this is likely the reason why 5 of the tools do not support it and assume that the input is a string, e.g., containing the full text of a report.
In summary, all \numtools tools can take an input string, read text from standard input or from a file, and extract indicators in the string.
In addition, three tools including our own, accept directly input HTML and PDF files, including the text extracted from the input document, before extracting indicators.
This is convenient for the user since it does not need to write its own text extraction code and allows the tool developer to customize the text extraction.

Text extraction is a common step in many Natural Language Processing (NLP) pipelines such as those used to analyze privacy policies (e.g.,~\cite{privee,slavin2016toward,zimmeck2017automated,polisis,policylint}).
Text extraction can have a significant impact on the extracted indicators. 
For example, the text extraction process may pre-pend or append text to the indicator value, making the extraction miss the indicator or output an incorrect indicator with extraneous characters.
It is also possible for the extracted text to split indicators across multiple lines, causing them to be missed.
For example, PDF text extraction libraries oftentimes split long URLs appearing in footnotes across multiple lines in the output text (e.g., URLs that do not fit the length of a column).

\updated{
  The only three tools that support all three input types are \Jager, \iocparser, and \searcher.
  These three tools 
}
use the pdfminer.six~\cite{pdfminer} library for PDF text extraction.
\Jager and \iocparser use the BeautifulSoup~\cite{bs4} library for HTML text extraction, while our \searcher tool supports both BeautifulSoup and Readability.js~\cite{readability}.
Recently, Hosseini et al.~\cite{hosseini21unifying} analyzed 7 HTML text extraction approaches used in privacy policy analysis showing wide variability among the produced text.
They concluded that the best-performing HTML text extraction libraries for privacy policies were Boilerpipe~\cite{boilerpipe} 
and Readability.js~\cite{readability}, while the worst-performing one was BeatifulSoup~\cite{bs4}.
BoilerPipe is written in Java and Readability.js in JavaScript, but both have Python wrappers available.

\subsection{Indicator Extraction}
\label{sec:extraction}
Given an input string, indicator extraction identifies indicators present in the string.
Seven of the tools use \regexps for the extraction, while \iocfinder uses grammars instead.
The tools using \regexps process the text in a loop, each time applying one \regexp at a time to the whole text. 
Each \regexp is associated with one indicator type that is assigned to each of the \regexp matches. 
Most tools have one \regexp for each supported indicator type, although some tools (e.g., \iocextract, \searcher) support multiple \regexps for the same indicator.

\juan{Describe grammar-based extraction}

The middle part of Table~\ref{tbl:tools} captures four properties of the indicator extraction step: 
whether defanged indicators are supported and rearmed, 
whether tools split the extraction process into matching and validation, 
whether the indicators are deduplicated, and
the supported indicator types.
We describe each property next.

\paragraph{Defanged and rearmed indicators.}
It is common for threat reports to \emph{defang} malicious indicators, (i.e., IOCs) in case a user inadvertently clicks on them in a navigation tool like a browser.
For example, IP address 9.9.9.9 may appear in a threat report as 9[.]9[.]9[.]9, while URL \url{http://example.com/badfile} may appear as \url{hxxp://example(.)com/badfile}.
Such indicators are often called \emph{defanged} to indicate that they have been converted into less harmful ones, and are not dangerous anymore.
These transformations are similar to the ones applied to email addresses to limit automated collection by spammers, e.g., ``contact\_at\_somewhere[.]com''.
Of the \numtools tools, only \jager does not identify defanged indicators. 
The other tools support a subset of the following defang transformations: 
(I) replacing the dot in IPv4 addresses, domain names, URLs, emails, and filenames,
(II) replacing the @ sign in emails,
(III) replacing the scheme in URLs (e.g., using hxxp:// instead of http://),
and (IV) replacing the backslash in URLs.
Two tools are marked as having partial support:
\iocparser only supports the dot in URLs and domain names, while
\cacador only supports the dot in IPv4 addresses.
The other tools are marked as full support.
However, they may not exactly support the same transformations, e.g., only \iocextractor supports the replacement of the backslash in URLs.
The space of defang transformations that users can apply is very large and the extraction tools only support the limited set presented above, which is clearly incomplete.
Intuitively, the defang transformations most important to support are those most often used by threat report authors (or by users when analyzing other sources like webpages).
However, it is hard to know the most popular transformations a priori, so it is typical to add support for new defang transformations as they are observed in the wild.

The tools use two approaches to handle defanged indicators.
The most popular approach, used by six tools, is broadening the \regexps used to identify the indicators to cover common defang operations.
For example, the \regexp should support that an IP address optionally contains brackets or parenthesis around the dots.
Once the defanged indicators have been matched by the \regexp, they can be \emph{rearmed} (or \emph{refanged}) to output the original indicator values.
Four tools rearm the defanged indicators by default, while another two (\iocparser, \iocextract) allow the user to choose if defanged or rearmed values should be returned. 
The alternative approach used by \iocfinder and \iocextractor is to first apply a rearm transformation to the raw text before applying the \regexps. 
This approach does not require broadening the \regexps to handle different defang transformations. 
On the other hand, blindly rearming the text before applying the \regexps could incorrectly modify the text (e.g., rearming some text that is not part of an indicator).
It also prevents returning the defanged value as it appears in the text since it has been rewritten.

\paragraph{Match and validate.}
One goal of indicator extraction is to minimize false positives (FPs), i.e., avoid outputting indicators that are not real indicators, such as two tokens concatenated with a period, that is not a fully qualified domain name.
This requires making the \regexps as narrow as possible, without introducing false negatives. 
For example, it is typical that a \regexp for domain names will check that the top-level domain (TLD) is one of the IANA-approved TLDs.
For this, it is common for tools to include a long list of valid TLDs inside the \regexp. 
Unfortunately, the IANA list already contains over 1,500 TLDs and may continue to grow over time making the \regexp cumbersome.
An alternative approach is to split the indicator extraction process into two steps: \regexp matching and validation. 
In this model, the \regexp can be a bit wider (i.e., produce more matches) because the validation, which is specific to each indicator type, will discard incorrect matches. 
This match and validate process was first used by \cyobstract, with basic checks such as validating that ASN numbers are in hard-coded ranges or that the length of a domain name is below 160 characters (which is not entirely correct as the maximum length is 255 octets).

Our \searcher tool implements a more complete match-and-validate process where each indicator type has an optional validation function that checks the returned \regexp matches. 
For example, in \searcher the \ioc{fqdn} \regexp does not need to include the list of valid TLDs because there is a \ioc{fqdn} validation function that ensures the matched value indeed contains a valid TLD. 
The advantage is that the list of IANA-approved TLDs can be kept in a separate file, which can be updated without modifying the \regexp.
More importantly, the match and validate approach allows to perform Turing-complete processing on the matches such as validating an embedded checksum in a bank account number or a Bitcoin address.
Such validation is not possible using only a \regexp, and it can significantly reduce FPs.
For example, it is common for MD5 hashes to match a Bitcoin \regexp, but it is highly unlikely those spurious matches will pass the checksum validation.
It is worth mentioning, that the implementation in our \searcher tool performs validation after the indicator has been rearmed so that the validation does not need to handle defang transformations.

\paragraph{Deduplication.}
It is possible for the same indicator to appear multiple times in the same string at different locations.
We say that a tool \textit{deduplicates} if it removes duplicated indicators from its output.
For example, if the same URL appears twice in the input string, the output only contains the \ioc{url} indicator once. 
Tools can be classified into those that always deduplicate (\CC), those that do not deduplicate (\cc), and those where deduplication is optional (\lc).
None of the prior tools return the position (i.e., start offset) at which the indicator is matched. 
Thus, the main value of not deduplicating is to know how many times an indicator appeared in the input string.
The most flexible approach is to make deduplication optional, as done by \iocparser and \iocfinder.

Our \searcher tool offers two different APIs. 
The raw API does not deduplicate.
It returns all indicators identified, with their indicator type, starting offset, raw value, and rearmed value. 
In contrast, the deduplicated API first invokes the raw API, and then it deduplicates the received values by removing the starting offset and the raw values, i.e., returns only deduplicated rearmed indicators.
Providing all matches with their starting offset allows the raw API to be used in additional scenarios. 
For example, Gao et al.~\cite{gao2021enabling} propose IOC protection to handle IOCs that contain dots (e.g., URLs, domains, IPs) in NLP pipelines.
Such indicators negatively impact sentence tokenization in NLP libraries that leverage dots to identify the end of sentences.
IOC protection first identifies the indicators (e.g., using \regexps), replaces their value in the text with a keyword that does not contain dots, tokenizes the text into sentences, and finally replaces back the keyword with the original indicator value. 
IOC protection requires the starting offset and the raw value of the indicator in the input string, precisely what our raw API provides.

\begin{table}[!ht]
\scriptsize
\centering
\setlength{\tabcolsep}{4pt}
\renewcommand{\arraystretch}{.8}
\caption{Indicators extracted by different tools.}
\label{tbl:iocsCompare}
\begin{tabular}{l|c|c|c|c|c|c|c|c|r}
\textbf{Indicator} & \featuretexta{\jager} & \featuretexta{\iocparser} & \featuretexta{\cacador} & \featuretexta{\cyobstract} & \featuretexta{\iocfinder} & \featuretexta{\iocextract} & \featuretexta{\iocextractor} & \featuretexta{\searcher} & \featuretexta{NumTools} \\
\hline
asn & & & & \Y & \Y & & \Y & & 3 \\ asnOwner & & & & \Y & & & & & 1 \\ attackType & & & & \Y & & & & & 1 \\ attCk & & & & & \Y & & & & 1 \\ authentihash & & & & & \Y & & & & 1 \\ avLabel & & & & \Y & & & & & 1 \\ bitcoin & & & & & \Y & & \Y & \Y & 3 \\ bitcoincash & & & & & & & & \Y & 1 \\ copyright & & & & & & & & \Y & 1 \\ country & & & & \Y & & & & & 1 \\ cve & \Y & \Y & \Y & \Y & \Y & & \Y & \Y & 7 \\ dashcoin & & & & & & & & \Y & 1 \\ dogecoin & & & & & & & & \Y & 1 \\ email & \Y & \Y & \Y & \Y & \Y & \Y & \Y & \Y & 8 \\ ethereum & & & & & & & \Y & \Y & 2 \\ facebookHandle & & & & & & & & \Y & 1 \\ filename & \Y & \Y & \Y & \Y & & & & & 4 \\ filepath & & \Y & & \Y & \Y & & & & 3 \\ fqdn & \Y & \Y & \Y & \Y & \Y & & \Y & \Y & 7 \\ githubHandle & & & & & & & & \Y & 1 \\ googleAdsense & & & & & \Y & & \Y & \Y & 3 \\ googleAnalytics & & & & & \Y & & \Y & \Y & 3 \\ googleTagManager & & & & & & & & \Y & 1 \\ iban & & & & & & & & \Y & 1 \\ icp & & & & & & & & \Y & 1 \\ importHash & & & & & \Y & & & & 1 \\ incident & & & & \Y & & & & & 1 \\ instagramHandle & & & & & & & & \Y & 1 \\ ip4 & \Y & \Y & \Y & \Y & \Y & \Y & \Y & \Y & 8 \\ ip4cidr & & & & \Y & \Y & & & \Y & 3 \\ ip4range & & & & \Y & & & & & 1 \\ ip6 & & & \Y & \Y & \Y & \Y & \Y & & 5 \\ ip6cidr & & & & \Y & & & & & 1 \\ ip6range & & & & \Y & & & & & 1 \\ isp & & & & \Y & & & & & 1 \\ linkedinHandle & & & & & & & & \Y & 1 \\ litecoin & & & & & & & & \Y & 1 \\ macAddress & & & & & \Y & & \Y & & 2 \\ md5 & \Y & \Y & \Y & \Y & \Y & \Y & \Y & \Y & 8 \\ monero & & & & & \Y & & \Y & \Y & 3 \\ onionAddress & & & & & & & & \Y & 1 \\ phoneNumber & & & & & \Y & & & \Y & 2 \\ pinterestHandle & & & & & & & & \Y & 1 \\ regKey & & \Y & & \Y & \Y & & & & 3 \\ sha1 & \Y & \Y & \Y & \Y & \Y & \Y & \Y & \Y & 8 \\ sha256 & \Y & \Y & \Y & \Y & \Y & \Y & \Y & \Y & 8 \\ sha512 & \Y & & \Y & & & \Y & \Y & & 4 \\ ssdeep & \Y & & \Y & \Y & \Y & & \Y & & 5 \\ telegramHandle & & & & & & & & \Y & 1 \\ tezos & & & & & & & & \Y & 1 \\ tlpLabel & & & & & \Y & & & & 1 \\ trademark & & & & & & & & \Y & 1 \\ twitterHandle & & & & & & & & \Y & 1 \\ url & \Y & \Y & \Y & \Y & \Y & \Y & \Y & \Y & 8 \\ userAgent & & & & \Y & \Y & & & & 2 \\ webmoney & & & & & & & & \Y & 1 \\ whatsappHandle & & & & & & & & \Y & 1 \\ xmppHandle & & & & & \Y & & & & 1 \\ yara & & & & & & \Y & & & 1 \\ youTubeChannel & & & & & & & & \Y & 1 \\ youtubeHandle & & & & & & & & \Y & 1 \\ zcash & & & & & & & & \Y & 1 \\ \hline
\end{tabular}
\end{table}

\paragraph{Indicators.}
\label{sec:tools-indicators}
One key difference between indicator extraction tools is the set of indicator
types they support, i.e., the set of indicators for which they have \regexps,
grammars, or extraction rules.
Overall, we have identified 63 indicators that the tools extract, as summarized in Table~\ref{tbl:iocsCompare}.
We group indicator types into 16 classes: 
network (13 indicators), 
social (12), 
blockchain (9),
cryptographic hashes (6),
analytics (3),
attack (3),
contact (2), 
file (2),
intellectual property (2),
organization (2), 
payment (2),
vulnerability (2),
fuzzy hash (1),
information sharing (1),
location (1), 
registry (1), and
yara (1). 

The largest class corresponds to network-related artifacts such as 
domain names (\ioc{fqdn}), 
URLs (\ioc{url}), 
IP addresses (\ioc{ip4}, \ioc{ip6}), 
IP subnets in CIDR form (\ioc{ip4cidr}, \ioc{ip6cidr}),
IP address ranges (\ioc{ipv4range}, \ioc{ipv6range}), 
AS numbers (\ioc{asn}), 
MAC addresses (\ioc{macaddress}),
Tor onion addresses used to access hidden services (\ioc{onionAddress}),
Internet Content Provider numbers that uniquely 
identify the owner of a Chinese website (\ioc{icp}), and 
HTTP User-Agent strings (\ioc{userAgent}).
The social category comprises of user handles for social networks 
(Facebook, Instagram, LinkedIn, Pinterest, Twitter), 
open source repositories (GitHub), 
Instant Messaging tools (Jabber, Telegram, WhatsApp), 
as well as YouTube usernames and channel identifiers.
The blockchain category comprises addresses for 9 popular blockchains
(Bitcoin, BitcoinCash, DashCoin, DogeCoin, Ethereum, LiteCoin, Monero, Tezos, ZCash).
Cryptographic hashes include MD5, SHA1, SHA256, and SHA512, as 
well as their application on specific parts of a Windows executable such as the import table (\ioc{importHash}) and the whole executable without Authenticode code-signing fields (\ioc{authentihash}).
There is also a fuzzy hash (\ioc{ssdeep}) used to identify files with similar content~\cite{ssdeep}.
The analytics category includes three Google identifiers:
Google Adsense, Google Analytics, and Google Tag Manager.
The attack category contains MITRE ATT\&CK techniques, tactics, and 
procedures, a list of attack-related keywords (e.g., DoS, spam), and 
antivirus labels (\ioc{avLabel}).
The payment category includes IBAN bank account numbers and WebMoney addresses~\cite{webmoney}.
The vulnerability category includes CVE identifiers and also a variety of identifiers for other vulnerability sources 
(e.g., BugTrack, Microsoft Bulletins) 
grouped into a generic \ioc{incident} indicator by \cyobstract.
The information-sharing category comprises a single indicator for the traffic
light protocol (\ioc{tlpLabel}) that controls the dissemination of
information~\cite{tlp}.
Indicators in the remaining categories are straightforward and capture: 
contact (\ioc{email}, \ioc{phoneNumber}), 
files (\ioc{filename}, \ioc{filepath}), 
intellectual property (\ioc{copyright}, \ioc{trademark}), 
organization names (\ioc{asnOwner}, \ioc{isp}),
locations (\ioc{country}), 
Windows registry keys (\ioc{registryKey}), and
Yara rules (\ioc{yara}).

There are only 6 indicators that are extracted by all \numtools tools: 
IPv4 addresses (\ioc{ip4}), emails, 
hashes (\ioc{md5}, \ioc{sha1}, \ioc{sha256}), and URLs.
Another two indicators are extracted by 7 tools: 
CVE vulnerability identifiers (\ioc{cve}) and domain names (\ioc{fqdn}).
Next come two indicators extracted by 5 tools:
IPv6 addresses (\ioc{ip6}) and SSDeep fuzzy hashes.
However, the majority corresponds to 38 (61\%) indicators extracted by a single tool.

Among the indicators in Table~\ref{tbl:iocsCompare} there are some that deserve discussion.
There are three indicators (\ioc{country}, \ioc{tlpLabel}, \ioc{attackType}) that are identified using \regexps that are a disjunction of keywords. 
For example, TLP labels can have only four values: \emph{TLP:white}, \emph{TLP:green}, \emph{TLP:amber}, \emph{TLP:red}.
Since the set of possible TLP values is finite, the \regexp can identify all possible indicator values. 
Similarly, there is only a finite number of recognized countries.
However, \ioc{attackType} includes a number of attack-related keywords that \cyobstract identifies. 
Such a list only includes attack topics of interest for the tool authors and may not include attack topics other users are interested in.
The approach of building a \regexp from a set of keywords can be applied to any taxonomy of terms (e.g., family names, exploit kit names).
\Cyobstract provides support for building such \regexps. 
The main limitation of this technique is that it cannot identify new terms 
(e.g., new family names or exploit kits).
\Cyobstract extracts two indicators that aim to capture organization names (\ioc{asnOwner}, \ioc{isp}).
However, organization names do not have a clear structure save when using company-related suffixes (e.g., Ltd., Inc.) and thus are typically extracted using Named Entity Recognition (NER) techniques based on machine learning classifiers.
Furthermore, trying to separate whether an organization name corresponds to an ISP or ASN owner is a challenging problem that is better suited for NLP techniques.
There are also some indicators that one could argue should be split into multiple indicators. 
For example, the \regexp for the \ioc{incident} indicator extracted by \cyobstract captures a disjunction of \regexps, each for a different vulnerability report identifier (e.g., BugTrack, Microsoft Bulletins). 
Such disjunction is more efficient than having a separate \regexp for each identifier but does not allow users to extract only those identifiers they are interested in. 
Another interesting case is cryptographic hashes like \ioc{authentihash} and \ioc{importHash} which are really \textit{subtypes} of other cryptographic hashes. 
For example, \ioc{importHash} is the MD5 of the import table of a PE executable~\cite{imphash}, while \ioc{authentihash} is the SHA256 of a PE executable excluding Authenticode code signing fields.
Thus, it is not possible for a \regexp to differentiate an \ioc{importHash} from a \ioc{md5} or an \ioc{authentihash} from a \ioc{sha256} except if the subtype indicators appear with some specific keywords before or after. 
Such \regexps are very specific, they rarely produce FPs but they will introduce FNs when the expected keywords do not appear before or following the hashes.
On the other hand, \regexps for the parent types will also identify indicators subtypes (e.g., the \ioc{md5} \regexp will also identify \ioc{importHash} indicators), but classifying an MD5 into what the MD5 captures (e.g., a file hash, a certificate hash, an import table hash) is a challenging problem that is may be easier to tackle through NLP.
There are also two indicators that correspond to ranges of IP addresses (\ioc{ip4range}, \ioc{ip6range}). 
Similar range indicators could be defined for any integer indicators, e.g., ASN numbers.
The last interesting case is Yara rules extracted by \iocextract.
Yara rules can be long and have an internal structure with mandatory and optional fields. 
Thus, some readers may not consider them indicators.

\subsection{Filtering.}
\label{sec:filtering}

 While most tools in Table~\ref{tbl:tools} label themselves as IOC extraction tools, in reality, what they extract are indicators.  
Not all indicators are IOCs; an indicator is an IOC only if it represents a 
malicious artifact.
For example, threat reports often contain hyperlinks to prior reports 
by other security vendors.
Those references may be extracted as \ioc{url} indicators,
but they can hardly be considered IOCs. 
We use the term \emph{generic indicators} to (emphatically) refer to indicators that are not IOCs, \ie not malicious.

Determining whether an indicator is benign or malicious, \ie whether 
it is an IOC, is challenging. 
Most tools in Table~\ref{tbl:tools} do not try to perform such determination and simply output all indicators they find.
But, three tools (\iocparser, \cacador, and our \searcher) implement a per-indicator filtering step, whose goal is to remove generic indicators leaving only the IOCs.
\iocparser and \cacador both make use of hard-coded blocklists (or \regexps) for benign indicators. 
Overall, the blocklists capture domains 
(or emails and URLs that contain those domains) 
from cybersecurity companies, news outlets, law enforcement agencies, and 
a few other trusted sites such as \url{github.com}.
\Cacador uses a single blocklist with 15 domains, 
12 of them from cybersecurity companies.
\Iocparser uses 11 blocklists, 
one per each indicator type it supports.
However, only 4 of those blocklists 
(\ioc{email}, \ioc{fqdn}, \ioc{ip4}, \ioc{url}) have entries.
The \ioc{email} blocklist contains 6 domains of cybersecurity companies.
The \ioc{fqdn} blocklist contains 143 domains of cybersecurity companies, 
news outlets, and law enforcement agencies.
The \ioc{url} blocklist contains 79 root URLs for popular cybersecurity blogs.
Finally, the \ioc{ip4} blocklist contains 6 reserved IP address ranges.

The main difference in our filtering module is that its
blocklist is dynamically generated, \ie created from the collected documents.
Indicators are added to the blocklist if they satisfy at least one of the
following 5 rules:
($i$) the indicator is a \ioc{fqdn}, \ioc{url}, or \ioc{email},  
whose domain is in the list of domains from where a document 
was collected;
($ii$) the indicator was extracted from at least 20 documents from the 
same origin; 
($iii$) the indicator is a \ioc{fqdn} or \ioc{url} whose domain 
(excluding the ``www.'' prefix if present) is 
in the top-100k of the Tranco domain popularity list~\cite{trancondss2019};
($iv$) the indicator appears in more than 90\% of all the collected documents; or
($v$) the indicator is a private IP address. 
The first two rules aim to filter indicators that belong to the origins from
where documents are collected, \eg email contact addresses for security
companies such as \ioc{contact@trendmicro.com}.  Instead of hard-coding domains
of cybersecurity companies and blogs, we infer them from the origins of the
collected documents. This allows each user to personalize the filtering to the
list of origins he chooses to monitor.
The next two rules target popular indicators that appear in most documents or 
in a list of popular domains. 
The frequently updated Tranco list is 
supplemented with a dynamically generated list of popular indicators 
appearing in the collected documents, 
which again can be personalized for each user.

Section~\ref{sec:evaluation} provides an evaluation of our 
filtering module and the filtering from \cacador and \iocparser 
on a manually generated ground truth.

\begin{table}[t]
  \scriptsize
  \centering
  \caption{Data collection summary.}
  \label{tbl:collection}
  \begin{tabular}{lrrrrr}
    \hline
    \multicolumn{4}{c}{}& \multicolumn{2}{c}{\textbf{Indicators}}\\
    \textbf{Source} & \textbf{Orig.} & \textbf{Start Date} & \textbf{Entries} & \textbf{Extracted} & \textbf{IOCs}\\
    \hline
    aptnotes   & \aptnotesOrigins   & 2022/07/18 &     629 &  44,540 &  41,031\\
    chainsmith & \chainsmithOrigins & 2022/07/18 &   3,792 &  48,046 &  41,246\\
    malpedia   & \malpediaOrigins   & 2022/07/18 &  11,363 & 271,495 & 237,011\\
    rss        & \rssOrigins        & 2021/04/01 &  97,882 & 314,351 & 210,149\\
    telegram   & \telegramOrigins   & 2021/09/09 &  61,892 &  44,534 &  31,445\\
    twitter    & \twitterOrigins    & 2021/10/28 & 397,333 & 338,967 & 126,081\\    
    \hline
    \textit{ALL} &  \textit{\numorigins} & 2021/04/01 & \textit{\numentries} & \textit{\numextractedindicators} & \textit{\numextractediocs}\\    
    \hline

  \end{tabular}
\end{table}

\section{Evaluation}
\label{sec:evaluation}

We have applied \tool to collect reports from six sources over the last 15
months.  We started our collection with RSS on April 1st, 2021.  We then added
Telegram on September 9th 2021, and Twitter on October 10th, 2021.
On July 18th, 2022, we added the three report datasets 
(\aptnotes, \chainsmith, \malpedia),
but these datasets are cumulative so \tool can also process their past entries. 

Table~\ref{tbl:collection} summarizes the data collection 
and IOC extraction results for each source. 
It captures the number of origins in the source at the end of the collection, 
the collection start date,
the number of collected entries, as well as
the number of extracted indicators (before filtering) and 
IOCs (after filtering).
The most diverse source by the number of origins is 
\malpedia with reports from \malpediaOrigins organizations, 
followed by \twitter with \twitterOrigins accounts, 
RSS with \rssOrigins feeds, and 
\aptnotes with reports from \aptnotesOrigins organizations.
In contrast, \chainsmith includes reports from only 11 blogs, and \tool 
tracks only \telegramOrigins Telegram channels.

\twitter is the source with the largest number of collected entries 
despite being introduced later than \rss and \telegram, 
but its entries are tweets and thus short compared to full reports 
collected from \rss and the three datasets.
\rss is the most stable contributor with 
an average of 6.1K new report URLs per month.
Among the three report datasets, 
\malpedia provides the most reports (11,363), with \chainsmith 
providing one-third (3,792), and \aptnotes only 629 reports. 
The relatively small size of \aptnotes is due to its more focused goal of only 
collecting reports for APTs.

\paragraph{IOC extraction.}
The last two columns in Table~\ref{tbl:collection} show the indicators
extracted by \searcher (before filtering) and the final IOCs (after filtering).
Overall, \tool extracted 978k indicators of which 618K (63\%) are IOCs. Thus,
filtering is extremely important for discarding over one-third of indicators
that are generic and thus are not IOCs.
Filtering affects each source differently.
In the extreme case of \twitter, two-thirds of the indicators are filtered.
Across all sources, domain names and IPv4 addresses 
account for 99\% of the generic indicators.
Filtered domain names are those matching the domain from where the report 
was collected and those included in the Tranco top-100K list. 
Filtered IPv4 addresses are those of private and local networks, 

The average number of IOCs per report is highest for the report datasets: 
\aptnotes (66), \malpedia (22), and \chainsmith (11).
Of those, \aptnotes and \malpedia are manually curated, 
which prevents the collection of non-technical reports. 
The collection in \chainsmith is automated, but it leverages a small 
number of blogs from large security companies known for 
their high-quality technical content.
In contrast, reports from \rss have a lower ratio of IOCs 
(2.2), likely due to the variety of feeds and reports they disseminate. 
For example, some feeds may focus on technology news for the wider public
(and thus provide fewer IOCs), 
while blogs from security companies may mix technical reports with 
less technical reports that focus on the virtues of their products.
Still, the large number of \rss feeds \tool monitors compensates for the 
lower ratio, making \rss a consistent IOC contributor.
Finally, \telegram (0.5) and \twitter (0.3) have the lowest ratios,
which is expected due to the limited content in each entry.

\begin{figure}[t]
  \centering
  \includegraphics[width=\columnwidth]{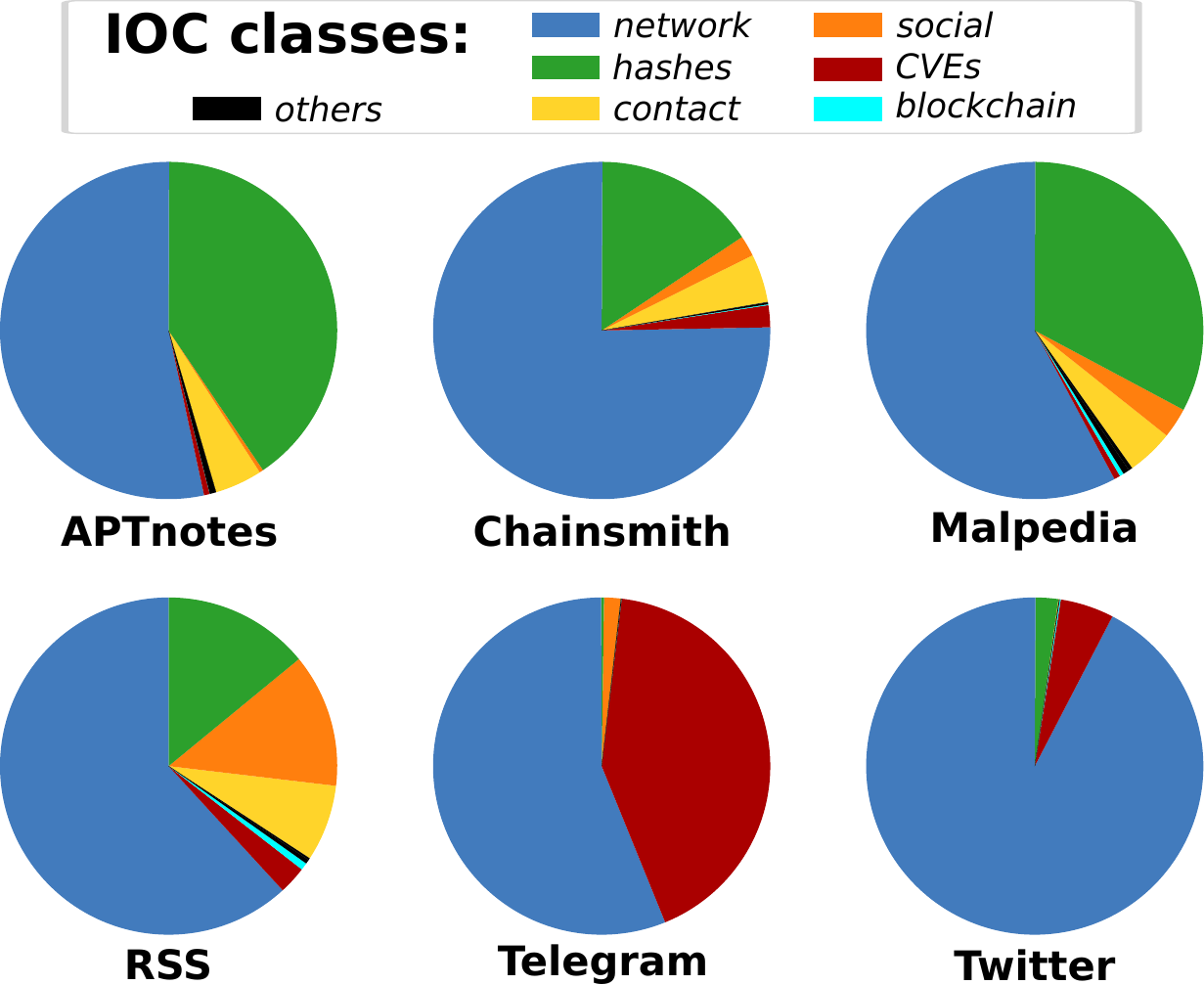}
  \caption{Distribution of IOC classes grouped per source.}
  \label{fig:ioc_classes_per_source}
\end{figure}

\paragraph{IOC class distribution.}
To understand what type of indicators each source distributes, 
we group indicators using the categories introduced 
in Section~\ref{sec:tools-indicators}.
Figure~\ref{fig:ioc_classes_per_source} reports the filtered IOCs 
split by class and source. 
The plots include only the six most popular classes, 
the rest are aggregated in the \textit{other} category. 
Across all sources, \textit{network} indicators dominate.
\malpedia, \telegram, and \twitter are the biggest contributors to 
the \textit{network} category, 
with more than 110K IOCs of this category per source.
The \textit{hash} category ranks second with most indicators 
in this category coming from \malpedia (82K file hash IOCs).
In third place we find \textit{contact} and \textit{social} indicators,
both having similar distributions with the vast majority coming 
from \rss and \malpedia.
These two sources are also the highest contributors of IOCs that 
survive the filtering and exhibit the highest variety in terms of 
indicator types.
For example, \rss and \malpedia provide 94\% of the \textit{blockchain} 
IOCs and over 61K email addresses, phone numbers and social network handles.
\telegram and \twitter show a significantly smaller variety of IOC types.
For example, the union of the \textit{contact} and \textit{social} 
classes represents only 1.6\% of the IOCs in \telegram, and 
they only account for 92 of the IOCs extracted from \twitter.
One possible reason is that \twitter accounts are more specialized with 
some accounts focusing only on certain indicator types such as malicious 
file hashes or vulnerability identifiers.

\begin{table*}[ht]
  \centering
  \setlength{\tabcolsep}{2pt}
  \caption{Top-10 origins that contribute with the highest number of IOCs, grouped per source. In brackets we report the percentage of IOCs associated to each origin.}
  \label{tbl:origins_contribution}
  \resizebox{\textwidth}{!}{  \begin{tabular}{r|rrrrrr}
    \hline
    \textbf{Rank} & \textbf{\aptnotes} & \textbf{\chainsmith} & \textbf{\malpedia} & \textbf{\rss} & \textbf{\telegram} & \textbf{\twitter}\\
    \hline
    \#1 &          Kaspersky (13.11\%) &                Webroot (27.92\%) & Palo Alto Networks ( 4.16\%) &   Dancho Danchev's Blog (26.21\%) &               cybsecurity (76.70\%) &        ecarlesi (76.42\%)\\
    \#2 & Palo Alto Networks ( 7.18\%) &                 Sucuri (22.07\%) &        Trend Micro ( 4.11\%) &    Blockchain on Medium ( 9.06\%) &         VulnerabilityNews ( 8.28\%) &     threatmeter ( 4.73\%)\\
    \#3 &             Norman ( 6.11\%) &           Malwarebytes (15.01\%) &          Kaspersky ( 2.92\%) & Cybersecurity on Medium ( 8.63\%) &  Cyber\_Security\_Channel ( 7.33\%) & malwrhunterteam ( 2.54\%)\\
    \#4 &           Symantec ( 5.24\%) &         Virus Bulletin (14.12\%) &               ESET ( 2.82\%) &             Cisco Talos ( 6.57\%) &                     cKure ( 4.01\%) &       bgpstream ( 2.02\%)\\
    \#5 &           ClearSky ( 4.74\%) &                 Sophos ( 4.88\%) &         Proofpoint ( 1.68\%) &   BleepingComputer News ( 2.44\%) &                     malwr ( 2.18\%) &   YourAnonRiots ( 1.54\%)\\
    \#6 &            FireEye ( 4.24\%) &             Forcepoint ( 4.60\%) &        Cisco Talos ( 1.66\%) &  Cointelegraph.com News ( 2.33\%) &            androidMalware ( 0.84\%) &   cryptolaemus1 ( 1.44\%)\\
    \#7 &               ESET ( 3.72\%) &                   ESET ( 4.34\%) &            FireEye ( 1.65\%) &                 F5 Labs ( 1.28\%) &                    ckuRED ( 0.48\%) &     ActorExpose ( 1.32\%)\\
    \#8 &        Trend Micro ( 2.67\%) &            TaoSecurity ( 4.15\%) &        BitDefender ( 1.44\%) &    Schneier on Security ( 1.26\%) &               canyoupwnme ( 0.16\%) &   MalwarePatrol ( 0.98\%)\\
    \#9 &        Citizen Lab ( 2.40\%) &               Hexacorn ( 1.70\%) &          360netlab ( 1.39\%) &       Malwarebytes Labs ( 1.12\%) &                itsecalert ( 0.03\%) &          1zrr4h ( 0.97\%)\\
    \#10&       PwC UK Blogs ( 1.88\%) &        Roger McClinton ( 0.90\%) &          Microsoft ( 1.34\%) &            contagiodump ( 1.08\%) &                                 -   &        dubstard ( 0.74\%)\\
  \end{tabular}
  }
\end{table*}

\paragraph{Top origins.}
Table~\ref{tbl:origins_contribution} shows the top-10 origins for each source, 
ranked by the percentage of IOCs the origin contributes to the source. 
Overall, we notice a long tail distribution, 
with the top-10 origins generating 50\% of all IOCs.
This pattern is consistent across all sources and IOC classes.
All three report datasets are dominated by large security companies, 
with the exception of the Citizen Lab at the University of Toronto 
(rank 8 in \aptnotes) and 
three personal blogs (ranks 8--10 in \chainsmith). 
A surprising contributor is Price Waterhouse Coopers 
(rank 10 in \aptnotes), one of the largest financial accounting companies, 
but less known for their security services.
In general, personal blogs exhibit less activity compared to company blogs
with 7\% of all IOCs coming from personal blogs.
\rss has more variety in the origins.
Surprisingly, the top contributor is the personal blog from Dancho Danchev, 
followed by two Medium blogs that aggregate 
blockchain and cybersecurity news.
The \rss top-10 is rounded by the research labs of three large companies
(Cisco, F5, Malwarebytes),  
two other personal blogs (Bruce Schneier, contagiodump), and 
two magazines (Cointelegraph.com and BleepingComputer).
Interestingly, two of the RSS top 10 origins focus on blockchain.
We added blockchain-specific sources motivated by
recent work that requires tagged Bitcoin addresses~\cite{btcrelations}.
\telegram is largely dominated by the \url{cybsecurity} channel, 
which provides 76\% of all Telegram IOCs.
Similarly, Twitter also has one dominant account \url{@ecarlesi} 
with more than 140K tweets, 76\% of all Twitter IOCs.
Second, but far behind with 4.7\% of IOCs, comes \url{@threatmeter}, 
an automated bot that publicizes vulnerabilities and
accounts for 90\% of \ioc{vulnerability} IOCs.

\begin{table*}[t]
\scriptsize
\centering
\setlength{\tabcolsep}{2.5pt}
\caption{Precision (P), recall (R), and F1 score achieved by each indicator 
extraction tool on the comparative evaluation over \numcomparisonreports 
documents. 
Only indicator types extracted by at least two tools are included.
}
\label{tbl:iocsEvalType}
\begin{tabular}{l|r|c|c|c|c|c|c|c|c|c|c|c|c|c|c|c|c|c|c|c|c|c|c|c|c}
\cline{3-26}
	\multicolumn{1}{l}{} & \multicolumn{1}{r|}{} & \multicolumn{3}{c|}{\textbf{\jager}} & \multicolumn{3}{c|}{\textbf{\iocparser}} & \multicolumn{3}{c|}{\textbf{\cacador}} & \multicolumn{3}{c|}{\textbf{\cyobstract}} & \multicolumn{3}{c|}{\textbf{\iocfinder}} & \multicolumn{3}{c|}{\textbf{\iocextract}} & \multicolumn{3}{c|}{\textbf{\iocextractor}} & \multicolumn{3}{c}{\textbf{\searcher}} \\
\hline
	\textbf{Indicator} & \textbf{Count} & \textbf{P} & \textbf{R} & \textbf{F1} & \textbf{P} & \textbf{R} & \textbf{F1} & \textbf{P} & \textbf{R} & \textbf{F1} & \textbf{P} & \textbf{R} & \textbf{F1} & \textbf{P} & \textbf{R} & \textbf{F1} & \textbf{P} & \textbf{R} & \textbf{F1} & \textbf{P} & \textbf{R} & \textbf{F1} & \textbf{P} & \textbf{R} & \textbf{F1} \\
\hline
	asn              & 202 & - & - & -  & - & - & -  & - & - & -  & 0.93 & 0.95 & 0.94  & \textbf{0.99} & \textbf{1.00} & \textbf{0.99}  & - & - & -  & 0.69 & 0.97 & 0.81  & - & - & - \\
	bitcoin          & 2,698 & - & - & -  & - & - & -  & - & - & -  & - & - & -  & 0.77 & 1.00 & 0.87  & - & - & -  & \textbf{1.00} & \textbf{1.00} & \textbf{1.00}  & 1.00 & 0.01 & 0.03 \\
	cve              & 2,135 & 0.96 & 0.99 & 0.98  & \textbf{1.00} & \textbf{1.00} & \textbf{1.00}  & - & - & -  & 0.98 & 0.97 & 0.97  & 0.98 & 1.00 & 0.99  & - & - & -  & 0.96 & 1.00 & 0.98  & \textbf{1.00} & \textbf{1.00} & \textbf{1.00} \\
	email            & 1,947 & 0.60 & 0.85 & 0.70  & 0.89 & 0.75 & 0.81  & 0.60 & 0.73 & 0.66  & 0.99 & 0.98 & 0.98  & 0.93 & 1.00 & 0.96  & 0.75 & 0.97 & 0.85  & 0.97 & 0.99 & 0.98  & \textbf{0.99} & \textbf{1.00} & \textbf{1.00} \\
	filename         & 17,082 & \textbf{0.98} & \textbf{0.97} & \textbf{0.98}  & 0.97 & 0.71 & 0.82  & 0.87 & 0.97 & 0.92  & 0.78 & 0.87 & 0.82  & - & - & -  & - & - & -  & - & - & -  & - & - & - \\
	filepath         & 1,551 & - & - & -  & \textbf{0.73} & \textbf{0.66} & \textbf{0.69}  & - & - & -  & 0.29 & 0.97 & 0.45  & 0.25 & 0.76 & 0.37  & - & - & -  & - & - & -  & - & - & - \\
	fqdn             & 41,360 & 0.48 & 0.06 & 0.10  & 0.99 & 0.91 & 0.94  & 0.56 & 0.39 & 0.46  & 0.97 & 0.96 & 0.97  & 0.95 & 0.99 & 0.97  & - & - & -  & 0.92 & 0.99 & 0.95  & \textbf{0.98} & \textbf{1.00} & \textbf{0.99} \\
	googleAdsense    & 4 & - & - & -  & - & - & -  & - & - & -  & - & - & -  & \textbf{1.00} & \textbf{1.00} & \textbf{1.00}  & - & - & -  & \textbf{1.00} & \textbf{1.00} & \textbf{1.00}  & 1.00 & 0.75 & 0.86 \\
	googleAnalytics  & 3 & - & - & -  & - & - & -  & - & - & -  & - & - & -  & \textbf{1.00} & \textbf{1.00} & \textbf{1.00}  & - & - & -  & 0.75 & 1.00 & 0.86  & \textbf{1.00} & \textbf{1.00} & \textbf{1.00} \\
	ip4              & 9,479 & 0.99 & 0.87 & 0.92  & 0.98 & 0.92 & 0.95  & 0.97 & 1.00 & 0.98  & 1.00 & 0.98 & 0.99  & 0.99 & 0.99 & 0.99  & 0.98 & 0.97 & 0.98  & 0.98 & 1.00 & 0.99  & \textbf{1.00} & \textbf{1.00} & \textbf{1.00} \\
	ip4cidr          & 287 & - & - & -  & - & - & -  & - & - & -  & \textbf{1.00} & \textbf{0.99} & \textbf{0.99}  & 0.97 & 1.00 & 0.99  & - & - & -  & - & - & -  & \textbf{1.00} & \textbf{0.99} & \textbf{0.99} \\
	ip6              & 967 & - & - & -  & - & - & -  & 0.50 & 0.91 & 0.65  & \textbf{0.87} & \textbf{0.66} & \textbf{0.75}  & 0.91 & 0.12 & 0.21  & 0.15 & 0.80 & 0.25  & 0.51 & 0.97 & 0.66  & - & - & - \\
	macAddress       & 64 & - & - & -  & - & - & -  & - & - & -  & - & - & -  & \textbf{1.00} & \textbf{1.00} & \textbf{1.00}  & - & - & -  & \textbf{1.00} & \textbf{1.00} & \textbf{1.00}  & - & - & - \\
	md5              & 14,635 & 1.00 & 0.98 & 0.99  & 1.00 & 1.00 & 1.00  & 0.45 & 1.00 & 0.62  & 1.00 & 0.97 & 0.98  & \textbf{1.00} & \textbf{1.00} & \textbf{1.00}  & 1.00 & 1.00 & 1.00  & \textbf{1.00} & \textbf{1.00} & \textbf{1.00}  & \textbf{1.00} & \textbf{1.00} & \textbf{1.00} \\
	monero           & 2 & - & - & -  & - & - & -  & - & - & -  & - & - & -  & \textbf{1.00} & \textbf{1.00} & \textbf{1.00}  & - & - & -  & \textbf{1.00} & \textbf{1.00} & \textbf{1.00}  & \textbf{1.00} & \textbf{1.00} & \textbf{1.00} \\
	regkey           & 608 & - & - & -  & 0.96 & 0.72 & 0.82  & - & - & -  & \textbf{0.80} & \textbf{0.90} & \textbf{0.85}  & 0.69 & 0.73 & 0.71  & - & - & -  & - & - & -  & - & - & - \\
	sha1             & 4,150 & \textbf{1.00} & \textbf{1.00} & \textbf{1.00}  & 1.00 & 0.99 & 0.99  & 0.41 & 1.00 & 0.58  & 1.00 & 0.99 & 0.99  & \textbf{1.00} & \textbf{1.00} & \textbf{1.00}  & 1.00 & 1.00 & 1.00  & \textbf{1.00} & \textbf{1.00} & \textbf{1.00}  & \textbf{1.00} & \textbf{1.00} & \textbf{1.00} \\
	sha256           & 5,336 & \textbf{1.00} & \textbf{1.00} & \textbf{1.00}  & \textbf{1.00} & \textbf{1.00} & \textbf{1.00}  & 0.97 & 1.00 & 0.99  & 1.00 & 0.94 & 0.97  & \textbf{1.00} & \textbf{1.00} & \textbf{1.00}  & \textbf{1.00} & \textbf{1.00} & \textbf{1.00}  & \textbf{1.00} & \textbf{1.00} & \textbf{1.00}  & \textbf{1.00} & \textbf{1.00} & \textbf{1.00} \\
	sha512           & 1 & \textbf{1.00} & \textbf{1.00} & \textbf{1.00}  & - & - & -  & 0.07 & 1.00 & 0.12  & - & - & -  & \textbf{1.00} & \textbf{1.00} & \textbf{1.00}  & \textbf{1.00} & \textbf{1.00} & \textbf{1.00}  & \textbf{1.00} & \textbf{1.00} & \textbf{1.00}  & - & - & - \\
	ssdeep           & 74 & 1.00 & 0.28 & 0.44  & - & - & -  & 0.16 & 0.30 & 0.21  & \textbf{0.81} & \textbf{0.88} & \textbf{0.84}  & 0.55 & 0.91 & 0.69  & - & - & -  & 0.47 & 1.00 & 0.64  & - & - & - \\
	url              & 14,818 & 0.61 & 0.56 & 0.59  & 0.53 & 0.56 & 0.54  & 0.51 & 0.52 & 0.52  & 0.76 & 0.96 & 0.85  & 0.62 & 0.83 & 0.71  & 0.60 & 0.99 & 0.75  & 0.70 & 0.80 & 0.74  & \textbf{0.77} & \textbf{1.00} & \textbf{0.87} \\
	\hline
	All            & 117,403 & 0.87 & 0.57 & 0.69 & 0.91 & 0.85 & 0.88 & 0.61 & 0.70 & 0.65 & 0.92 & 0.96 & 0.94 & 0.87 & 0.96 & 0.91 & 0.77 & 0.99 & 0.87 & 0.90 & 0.96 & 0.93 & \textbf{0.95} & \textbf{0.97} & \textbf{0.96} \\

	\ignore{
	asn              & 74 & - & - & -  & - & - & -  & - & - & -  & 0.92 & 0.95 & 0.93  & \textbf{1.00} & \textbf{1.00} & \textbf{1.00}  & - & - & -  & 0.66 & 0.95 & 0.78  & - & - & - \\
	bitcoin          & 1,365 & - & - & -  & - & - & -  & - & - & -  & - & - & -  & 0.98 & 1.00 & 0.99  & - & - & -  & \textbf{1.00} & \textbf{1.00} & \textbf{1.00}  & 1.00 & 0.00 & 0.01 \\
	cve              & 593 & 0.91 & 0.99 & 0.95  & \textbf{1.00} & \textbf{1.00} & \textbf{1.00}  & - & - & -  & 0.95 & 0.96 & 0.96  & 0.95 & 1.00 & 0.98  & - & - & -  & 0.91 & 1.00 & 0.95  & \textbf{1.00} & \textbf{1.00} & \textbf{1.00} \\
	email            & 991 & 0.71 & 0.81 & 0.76  & 0.88 & 0.68 & 0.76  & 0.65 & 0.73 & 0.69  & 0.99 & 0.96 & 0.98  & 0.95 & 1.00 & 0.97  & 0.85 & 0.99 & 0.91  & 0.97 & 0.99 & 0.98  & \textbf{0.99} & \textbf{1.00} & \textbf{0.99} \\
	filename         & 8,015 & \textbf{0.99} & \textbf{0.95} & \textbf{0.97}  & 0.98 & 0.91 & 0.94  & 0.89 & 0.98 & 0.93  & 0.75 & 0.86 & 0.80  & - & - & -  & - & - & -  & - & - & -  & - & - & - \\
	filepath         & 947 & - & - & -  & \textbf{0.71} & \textbf{0.67} & \textbf{0.69}  & - & - & -  & 0.25 & 0.97 & 0.40  & 0.37 & 0.74 & 0.49  & - & - & -  & - & - & -  & - & - & - \\
	fqdn             & 17,139 & 0.52 & 0.12 & 0.19  & 0.98 & 0.92 & 0.95  & 0.47 & 0.27 & 0.34  & 0.97 & 0.92 & 0.95  & 0.96 & 1.00 & 0.98  & - & - & -  & 0.93 & 0.99 & 0.96  & \textbf{0.98} & \textbf{1.00} & \textbf{0.99} \\
	ip4              & 4,356 & 0.99 & 0.82 & 0.90  & 0.99 & 0.86 & 0.92  & 0.97 & 1.00 & 0.98  & 1.00 & 0.97 & 0.99  & 0.99 & 0.99 & 0.99  & 1.00 & 0.98 & 0.99  & 0.99 & 1.00 & 1.00  & \textbf{1.00} & \textbf{1.00} & \textbf{1.00} \\
	ip4cidr          & 118 & - & - & -  & - & - & -  & - & - & -  & 1.00 & 0.97 & 0.99  & 0.98 & 1.00 & 0.99  & - & - & -  & - & - & -  & \textbf{1.00} & \textbf{1.00} & \textbf{1.00} \\
	ip6              & 658 & - & - & -  & - & - & -  & 0.62 & 0.97 & 0.76  & 0.88 & 0.52 & 0.65  & 1.00 & 0.04 & 0.08  & 0.17 & 0.82 & 0.28  & \textbf{0.63} & \textbf{1.00} & \textbf{0.77}  & - & - & - \\
	macAddress       & 11 & - & - & -  & - & - & -  & - & - & -  & - & - & -  & \textbf{1.00} & \textbf{1.00} & \textbf{1.00}  & - & - & -  & \textbf{1.00} & \textbf{1.00} & \textbf{1.00}  & - & - & - \\
	md5              & 10,267 & 1.00 & 0.97 & 0.99  & 1.00 & 0.99 & 1.00  & 0.46 & 1.00 & 0.63  & 1.00 & 0.95 & 0.98  & \textbf{1.00} & \textbf{1.00} & \textbf{1.00}  & \textbf{1.00} & \textbf{1.00} & \textbf{1.00}  & \textbf{1.00} & \textbf{1.00} & \textbf{1.00}  & \textbf{1.00} & \textbf{1.00} & \textbf{1.00} \\
	regkey           & 430 & - & - & -  & 0.97 & 0.71 & 0.82  & - & - & -  & \textbf{0.83} & \textbf{0.87} & \textbf{0.85}  & 0.75 & 0.74 & 0.75  & - & - & -  & - & - & -  & - & - & - \\
	sha1             & 3,036 & \textbf{1.00} & \textbf{1.00} & \textbf{1.00}  & 1.00 & 0.99 & 0.99  & 0.41 & 1.00 & 0.59  & 1.00 & 0.99 & 0.99  & \textbf{1.00} & \textbf{1.00} & \textbf{1.00}  & \textbf{1.00} & \textbf{1.00} & \textbf{1.00}  & \textbf{1.00} & \textbf{1.00} & \textbf{1.00}  & \textbf{1.00} & \textbf{1.00} & \textbf{1.00} \\
	sha256           & 4,052 & \textbf{1.00} & \textbf{1.00} & \textbf{1.00}  & \textbf{1.00} & \textbf{1.00} & \textbf{1.00}  & 0.98 & 1.00 & 0.99  & 1.00 & 0.92 & 0.96  & \textbf{1.00} & \textbf{1.00} & \textbf{1.00}  & \textbf{1.00} & \textbf{1.00} & \textbf{1.00}  & \textbf{1.00} & \textbf{1.00} & \textbf{1.00}  & \textbf{1.00} & \textbf{1.00} & \textbf{1.00} \\
	sha512           & 1 & \textbf{1.00} & \textbf{1.00} & \textbf{1.00}  & - & - & -  & 0.09 & 1.00 & 0.17  & - & - & -  & \textbf{1.00} & \textbf{1.00} & \textbf{1.00}  & \textbf{1.00} & \textbf{1.00} & \textbf{1.00}  & \textbf{1.00} & \textbf{1.00} & \textbf{1.00}  & - & - & - \\
	ssdeep           & 43 & 1.00 & 0.19 & 0.31  & - & - & -  & 0.16 & 0.19 & 0.17  & \textbf{0.98} & \textbf{0.93} & \textbf{0.95}  & 0.84 & 1.00 & 0.91  & - & - & -  & 0.84 & 1.00 & 0.91  & - & - & - \\
	url              & 5,195 & 0.70 & 0.70 & 0.70  & 0.52 & 0.56 & 0.54  & 0.48 & 0.51 & 0.49  & 0.78 & 0.93 & 0.85  & 0.67 & 0.90 & 0.77  & 0.45 & 1.00 & 0.62  & 0.80 & 0.90 & 0.85  & \textbf{0.78} & \textbf{1.00} & \textbf{0.87} \\
	\hline
	All              & 57,291 & 0.90 & 0.66 & 0.76 & 0.93 & 0.89 & 0.91 & 0.59 & 0.71 & 0.65 & 0.90 & 0.94 & 0.92 & 0.92 & 0.97 & 0.94 & 0.76 & 0.99 & 0.86 & 0.94 & 0.99 & 0.96 & \textbf{0.96} & \textbf{0.97} & \textbf{0.97} \\
	}
	\ignore{
	asn              & 128 & - & - & -  & - & - & -  & - & - & -  & 0.94 & 0.95 & 0.94  & \textbf{0.98} & \textbf{0.99} & \textbf{0.99}  & - & - & -  & 0.71 & 0.98 & 0.82  & - & - & - \\
	bitcoin          & 1,333 & - & - & -  & - & - & -  & - & - & -  & - & - & -  & 0.64 & 1.00 & 0.78  & - & - & -  & \textbf{1.00} & \textbf{0.99} & \textbf{1.00}  & 1.00 & 0.02 & 0.05 \\
	cve              & 1,542 & 0.99 & 1.00 & 0.99  & \textbf{1.00} & \textbf{1.00} & \textbf{1.00}  & - & - & -  & 0.99 & 0.97 & 0.98  & 1.00 & 1.00 & 1.00  & - & - & -  & 0.99 & 1.00 & 0.99  & \textbf{1.00} & \textbf{1.00} & \textbf{1.00} \\
	email            & 956 & 0.53 & 0.88 & 0.66  & 0.90 & 0.83 & 0.86  & 0.56 & 0.73 & 0.63  & 0.99 & 0.99 & 0.99  & 0.92 & 0.99 & 0.96  & 0.67 & 0.95 & 0.79  & 0.96 & 0.99 & 0.97  & \textbf{1.00} & \textbf{1.00} & \textbf{1.00} \\
	filename         & 9,067 & \textbf{0.98} & \textbf{0.98} & \textbf{0.98}  & 0.97 & 0.53 & 0.69  & 0.86 & 0.96 & 0.90  & 0.80 & 0.89 & 0.84  & - & - & -  & - & - & -  & - & - & -  & - & - & - \\
	filepath         & 604 & - & - & -  & \textbf{0.75} & \textbf{0.63} & \textbf{0.68}  & - & - & -  & 0.39 & 0.98 & 0.56  & 0.17 & 0.79 & 0.28  & - & - & -  & - & - & -  & - & - & - \\
	fqdn             & 24,221 & 0.32 & 0.01 & 0.02  & 0.99 & 0.90 & 0.94  & 0.61 & 0.48 & 0.54  & 0.98 & 0.98 & 0.98  & 0.94 & 0.99 & 0.96  & - & - & -  & 0.91 & 0.98 & 0.95  & \textbf{0.98} & \textbf{1.00} & \textbf{0.99} \\
	googleAdsense    & 4 & - & - & -  & - & - & -  & - & - & -  & - & - & -  & \textbf{1.00} & \textbf{1.00} & \textbf{1.00}  & - & - & -  & \textbf{1.00} & \textbf{1.00} & \textbf{1.00}  & 1.00 & 0.75 & 0.86 \\
	googleAnalytics  & 4 & - & - & -  & - & - & -  & - & - & -  & - & - & -  & 1.00 & 0.75 & 0.86  & - & - & -  & \textbf{0.80} & \textbf{1.00} & \textbf{0.89}  & 0.33 & 0.25 & 0.29 \\
	ip4              & 5,123 & 0.98 & 0.91 & 0.94  & 0.97 & 0.98 & 0.98  & 0.97 & 1.00 & 0.98  & 1.00 & 1.00 & 1.00  & 0.99 & 0.98 & 0.98  & 0.97 & 0.97 & 0.97  & 0.97 & 1.00 & 0.98  & \textbf{1.00} & \textbf{1.00} & \textbf{1.00} \\
	ip4cidr          & 169 & - & - & -  & - & - & -  & - & - & -  & \textbf{1.00} & \textbf{1.00} & \textbf{1.00}  & 0.97 & 1.00 & 0.99  & - & - & -  & - & - & -  & 1.00 & 0.98 & 0.99 \\
	ip6              & 309 & - & - & -  & - & - & -  & 0.33 & 0.79 & 0.46  & \textbf{0.86} & \textbf{0.97} & \textbf{0.91}  & 0.89 & 0.28 & 0.43  & 0.11 & 0.76 & 0.20  & 0.34 & 0.90 & 0.50  & - & - & - \\
	macAddress       & 53 & - & - & -  & - & - & -  & - & - & -  & - & - & -  & \textbf{1.00} & \textbf{1.00} & \textbf{1.00}  & - & - & -  & \textbf{1.00} & \textbf{1.00} & \textbf{1.00}  & - & - & - \\
	md5              & 4,368 & \textbf{1.00} & \textbf{1.00} & \textbf{1.00}  & \textbf{1.00} & \textbf{1.00} & \textbf{1.00}  & 0.43 & 1.00 & 0.60  & \textbf{1.00} & \textbf{1.00} & \textbf{1.00}  & 0.99 & 1.00 & 1.00  & 0.99 & 1.00 & 0.99  & \textbf{1.00} & \textbf{1.00} & \textbf{1.00}  & \textbf{1.00} & \textbf{1.00} & \textbf{1.00} \\
	monero           & 2 & - & - & -  & - & - & -  & - & - & -  & - & - & -  & \textbf{1.00} & \textbf{1.00} & \textbf{1.00}  & - & - & -  & \textbf{1.00} & \textbf{1.00} & \textbf{1.00}  & \textbf{1.00} & \textbf{1.00} & \textbf{1.00} \\
	regkey           & 178 & - & - & -  & 0.96 & 0.74 & 0.83  & - & - & -  & \textbf{0.74} & \textbf{0.97} & \textbf{0.84}  & 0.58 & 0.70 & 0.63  & - & - & -  & - & - & -  & - & - & - \\
	sha1             & 1,114 & \textbf{1.00} & \textbf{1.00} & \textbf{1.00}  & \textbf{1.00} & \textbf{1.00} & \textbf{1.00}  & 0.41 & 1.00 & 0.58  & 1.00 & 0.99 & 1.00  & \textbf{1.00} & \textbf{1.00} & \textbf{1.00}  & 0.98 & 1.00 & 0.99  & \textbf{1.00} & \textbf{1.00} & \textbf{1.00}  & \textbf{1.00} & \textbf{1.00} & \textbf{1.00} \\
	sha256           & 1,284 & 1.00 & 0.99 & 1.00  & 1.00 & 0.99 & 1.00  & 0.95 & 1.00 & 0.98  & 1.00 & 0.99 & 1.00  & 0.99 & 1.00 & 1.00  & 0.99 & 1.00 & 1.00  & \textbf{1.00} & \textbf{1.00} & \textbf{1.00}  & \textbf{1.00} & \textbf{1.00} & \textbf{1.00} \\
	ssdeep           & 31 & 1.00 & 0.42 & 0.59  & - & - & -  & 0.16 & 0.45 & 0.24  & \textbf{0.64} & \textbf{0.81} & \textbf{0.71}  & 0.34 & 0.77 & 0.48  & - & - & -  & 0.29 & 1.00 & 0.45  & - & - & - \\
	url              & 9,623 & 0.56 & 0.49 & 0.52  & 0.53 & 0.55 & 0.54  & 0.53 & 0.53 & 0.53  & 0.76 & 0.98 & 0.86  & 0.59 & 0.79 & 0.67  & 0.72 & 0.99 & 0.84  & 0.64 & 0.74 & 0.69  & \textbf{0.76} & \textbf{1.00} & \textbf{0.86} \\
	\hline
	All              & 60,113 & 0.84 & 0.48 & 0.61 & 0.89 & 0.80 & 0.84 & 0.63 & 0.68 & 0.66 & \textbf{0.93} & \textbf{0.98} & \textbf{0.96} & 0.82 & 0.95 & 0.88 & 0.79 & 0.98 & 0.87 & 0.87 & 0.94 & 0.90 & \textbf{0.94} & \textbf{0.97} & \textbf{0.96} \\
	}
\hline
\end{tabular}
\end{table*}

\section{Evaluation of IOC Extraction Tools}
\label{sec:evalTools}

To evaluate the indicator extraction tools, we design a majority-vote 
methodology that runs the \numtools tools in Table~\ref{tbl:tools} on the text of the same set of threat reports.
Then, we compare the indicators extracted by the different tools on the same report, assuming that the correct indicators are those extracted by a 
\textit{majority} of tools.
The advantage of such evaluation is that it can be run on a large set of reports bypassing the challenge of building a ground truth that is representative of the wealth of indicators the tools extract.
The disadvantage is that this approach works only when indicators are extracted by multiple tools, and in some occasions it is possible that the \textit{minority} of tools is actually correct.
We detail our methodology and the obtained results next.

We first build a document dataset by extracting the text using \tool from all reports in \aptnotes and \chainsmith.
We choose these two sources because they cover different report file types
(\aptnotes has PDF reports and \chainsmith mostly HTML reports) and 
because both sources have a high average of IOCs per report, 
as shown in Section~\ref{sec:evaluation}.
In four PDF reports the text extraction failed:
two were Excel spreadsheets that our text extraction does not support and 
the other two were corrupted.
We evaluate the tools using the \numcomparisonreports successfully extracted text documents.

\begin{algorithm}[t]
\footnotesize
\caption{Accuracy comparison methodology.}
\label{alg:comparison}
\begin{algorithmic}[1]
	\Procedure{compare}{$tools$, $docs$}
	\State $accuracy \gets \{\}$

    \For{tool \textbf{in} tools}
      \State $accuracy[tool] \gets (0,0,0,0)$ 
	\EndFor

  	\For{doc \textbf{in} docs}
    \State $extracted \gets \{\}$
    \State $all \gets set()$
    \For{tool \textbf{in} tools}
      \State $iocs \gets Run(tool, doc)$
      \State $iocs \gets Normalize(iocs)$
      \State $all \gets all \cup iocs$
      \For{ioc \textbf{in} iocs}
        \State $extracted[ioc].add(tool)$
      \EndFor
    \EndFor
    \For{ioc \textbf{in} all}
      \State $found \gets extracted[ioc]$ 
      \State $supported \gets canExtract(ioc.type)$ 
      \State $missed \gets supported - found$ 
      \If{len(found) \textgreater \, len(missed)}
          \For{tool \textbf{in} found}
            \State $accuracy[tool].tp \mathrel{+}= 1$ 
          \EndFor
          \For{tool \textbf{in} missed}
            \State $accuracy[tool].fn \mathrel{+}= 1$
          \EndFor
      \EndIf
      \If{len(missed) \textgreater \, len(found)}
          \For{tool \textbf{in} found}
            \State $accuracy[tool].fp \mathrel{+}= 1$
          \EndFor
          \For{tool \textbf{in} missed}
            \State $accuracy[tool].tn \mathrel{+}= 1$
          \EndFor
      \EndIf
    \EndFor
	\EndFor
	\State \textbf{return} $accuracy$
	\EndProcedure
\end{algorithmic}
\end{algorithm}

\paragraph{Methodology.}
Algorithm~\ref{alg:comparison} details our 
accuracy comparison methodology.
We run all tools on each document, saving the indicators each tool extracts to a separate file.
To compare all tools in a similar setting, we disable filtering for tools that support it (\iocparser, \cacador, \searcher) and deduplicate indicators.
Since each tool may assign slightly different names to indicator types (e.g. ipv4addr, ip, and ipv4 for IPv4 addresses), we normalize them to match the names in Table~\ref{tbl:iocsCompare}.
Additionally, some tools transform case-insensitive indicator values (e.g., domain names, email addresses) to lowercase or uppercase, while others output them as they appear in the text.
To address such differences, we also normalize indicator values.
In particular, we lowercase the following indicator values: hashes (\ioc{md5}, \ioc{sha1}, \ioc{sha256}, \ioc{sha512}, \ioc{ssdeep}), \ioc{regkey}, \ioc{ip6}, \ioc{fqdn}, and \ioc{email}.
Finally, we normalize AS numbers to the format \textit{AS1234} (e.g. from asn1234 to AS1234) and URLs by prepending ``http://'' when no scheme is present.

For each tool, the evaluation keeps counters for true positives (TPs), false
positives (FPs), false negatives (FNs), and true negatives (TNs).
The evaluation processes one document at a time, updating the counters with the document results.
For each document, it examines all indicators extracted from the document by at least one tool.
For each indicator, it generates three sets: 
the \emph{found} set captures the tools that identified the indicator in the document, 
the \emph{missed} set captures the tools that support the indicator type but did not identify it, and
the \emph{unsupported} set captures the tools that do not support the indicator type.
If the size of the found set is larger than the size of the missed set, then the evaluation assumes the majority is correct and therefore the indicator was indeed present in the document. 
Thus, it adds a TP for each tool in the found set and a FN for each tool in the missed set.
If the size of the missed set is larger than the size of the found set, then the evaluation assumes the majority is correct and therefore the indicator was not present in the document. 
Thus, it adds a FP for each tool in the found set and a TN for each tool in the missed set. 
If the size of the found and missed sets is the same, then there is no majority. 
In such cases, we skip the indicator and do not update the counters.
In future work, we plan to investigate how to break such ties.
After processing all documents, we compute the precision, recall, and F1 score for each tool across all indicators, as well as separately for each indicator type. 

\paragraph{Handling errors.}
When running all tools on the \numcomparisonreports documents we observed that \iocextract, the second most popular tool, needed hours to process some documents, compared to seconds or a few minutes for other tools.
We suspected that one of their \regexps had a ReDoS vulnerability, which caused a worst case that got triggered only in some documents~\cite{davis2018impact}. 
We found an open issue in the \iocextract repository regarding catastrophic backtracking in the \regexp used to identify defanged URLs that modify the backslash~\cite{iocextractRedos}.
To be able to complete the evaluation in a reasonable time, we configured \iocextract to avoid using the problematic \regexp. 
This change does not affect the accuracy results, as \iocextract is the only tool supporting that defang transformation.
Similarly, there are two documents where \iocextractor does not terminate.
We identify the root of this issue in the \regexps used to extract domain names.
Of the four domain \regexps used, only one avoids the problem, while the others require hours to process the document.
For these two documents, we added \iocextractor to the missed set for all indicators in the document.
These results show that ReDoS vulnerabilities are a serious problem for 
indicator extraction tools, but oftentimes these worst cases appear only
when examining a large number of documents.
We examine whether ReDoS vulnerabilities exist in the \regexps 
used by \searcher with the \emph{rat} ReDoS checker~\cite{ratRedos}.
The tool implements a sound static analysis (\ie no false positives) that 
identifies \regexps with exponential time worst case~\cite{parolini2022sound}. 
Except for 3 \regexps the \emph{rat} tool could not handle due to negative lookbehinds,
all other \searcher \regexps are free from exponential time worst case.

There are also a few documents where a tool throws an exception and 
thus extracts no indicators. 
This happens for \cyobstract in \numexcyobstract documents and 
for \jager in \numexjager.
For \cyobstract, 
only the indicators of the type causing the exception are missed.
We included \jager in the missed set for all indicators in those documents.

\paragraph{Results.}
Table~\ref{tbl:iocsEvalType} summarizes the results for the 21 indicator types that are extracted by more than one tool. 
The columns show 
the indicator type,
the number of indicators of that type considered TPs,
as well as the precision (P), recall (R), and F1 score for each tool.
We do not include indicators extracted only by a single tool, 
as there is no concept of majority for those.

Overall, \searcher is the tool that achieves both 
the best precision (0.95) and F1 score (0.96),
while \iocextract achieves the best recall (0.99).
The highest overall F1 scores are for 
\searcher (0.96), \cyobstract (0.94), \iocextractor (0.93), and 
\iocfinder (0.91).
The lowest F1 scores are for \cacador (0.65) and \jager (0.69).
These results seem to indicate that recent tools 
(\searcher, \iocextractor, \iocfinder) 
perform better than those released earlier (\jager, \cacador), 
possibly because newer tools could use older ones for comparison.

The table also presents the accuracy results for each indicator type. 
We observe perfect agreement for MAC address extraction
(although only 64 MAC addresses are found by two tools) and nearly perfect 
agreement on the hashes, with the exception of \cacador which has 
low F1 scores for \ioc{sha512} (0.12), \ioc{sha1} (0.58), and \ioc{md5} (0.62).
\Searcher performs best in 11 of the 13 (85\%) indicator types it supports in the table, 
\iocextractor in 8 out of 17 (47\%), and 
\iocfinder in 9 out of 20 (45\%).
\Cyobstract is the best tool for extracting registry keys (0.85) and 
SSDeep hashes (0.84);
\jager for extracting filenames (0.98), 
\iocparser for filepaths (0.69), and 
\searcher for emails (1.0), domain names (0.99), IPv4 addresses (1.0), and 
URLs (0.87).
These results are useful for future work in this area, to identify which prior tool may have the best \regexp for an indicator type. 
For example, if we were to add registry key support to \searcher we would start by looking at \cyobstract and for filenames to \jager.

The indicator types with the lowest agreement are 
IPv6 addresses with F1 score ranging 0.21--0.75; 
filepaths (0.37--0.69); 
URLs (0.52--0.87); and
registry keys (0.71--0.85).
These are arguably the indicator types for which \regexps are harder to build.
We observe that most IPv6 addresses extracted are actually FPs caused by a
\regexp that matches serial numbers in certificates and certificate
fingerprints.

One surprising result is that \searcher has an F1 score of 0.03 on Bitcoin
addresses, while it achieves the best F1 score on most other indicators it
extracts.
We manually checked the extracted Bitcoin addresses and observed that \searcher
is actually always correct in identifying Bitcoin addresses.
Both \iocextractor and \iocfinder return hashes that are erroneously included as Bitcoin addresses. 
\Searcher avoids those FPs by examining the checksum in the \regexp matches, which do not validate.
This is an example of the minority of tools being correct and the majority being wrong. 
We discuss this issue in Section~\ref{sec:discussion}.

\begin{table}[t]
\scriptsize
\centering
\caption{Top-10 indicator types extracted by all tools.}
\label{tbl:top10}
\begin{tabular}{lrr}
\hline
	\textbf{Indicator} & \textbf{Count} & \textbf{Tools} \\
\hline
	fqdn             & 41,360 & 7 \\
	attackType       & 26,109 & 1 \\
	country          & 17,654 & 1 \\
	filename         & 17,082 & 4 \\
	url              & 14,818 & 8 \\
	md5              & 14,635 & 8 \\
	ip4              & 9,479 & 8 \\
	sha256           & 5,336 & 8 \\
	avLabel          & 4,660 & 1 \\
	sha1             & 4,150 & 8 \\
\hline
\end{tabular}
\end{table}

Table~\ref{tbl:top10} shows the Top 10 most popular indicator types across the \numcomparisonreports documents.
The count column represents the number of TPs, and the last column reports the number of tools that extract each indicator.
Among those, 7 correspond to indicators supported by most tools, namely domain names, file names, URLs, IPv4 addresses, and hashes (\ioc{md5}, \ioc{sha256}, \ioc{sha1}).
On the other hand, the other three indicators are extracted only by \cyobstract and correspond to attack-type keywords, countries, and AV labels.
A total of 16 indicator types are not found in the \numcomparisonreports 
reports.
These include 
\ioc{attribution}, \ioc{groupName}, \ioc{authentihash}, \ioc{ip6range}, \ioc{useragent}, \ioc{iban}, \ioc{icp}, 7 blockchain addresses 
(\ioc{dashcoin}, \ioc{dogecoin}, \ioc{ethereum}, \ioc{litecoin}, 
\ioc{tezos}, \ioc{webmoney}, \ioc{zcash}), and 
two social handles (\ioc{telegramHandle}, \ioc{whatsappHandle}).
These correspond to less popular indicators in \aptnotes and \chainsmith 
reports.
However, other sources may differ. 
For example, \rss collection includes blockchain-related blogs, 
which leads to \searcher extracting indicators for all blockchain 
addresses in Section~\ref{sec:evaluation} 
(e.g., 726 \ioc{ethereum} addresses).

\begin{table}[t]
  \scriptsize
  \centering
  \caption{Post-hoc t-tests between groups summary.}
  \label{tbl:anova}
  \begin{tabular}{llrrr}
  \hline
	\textbf{A} & \textbf{B} & \textbf{p-unc} & \textbf{p-corr} & \textbf{eta-square} \\
  \hline
\cacador	 & \cyobstract	 & 7.6527e-03	 & 1.0203e-02	 & 3.6639e-04 \\
\cacador	 & \iocextract	 & 3.1698e-21	 & 1.4792e-20	 & 1.2091e-03 \\
\cacador	 & \iocextractor	 & 2.9591e-18	 & 1.1836e-17	 & 8.6447e-04 \\
\cacador	 & \iocfinder	 & 1.12366-17	 & 3.9328e-17	 & 8.48226-04 \\
\cacador	 & \iocparser	 & 7.11888-05	 & 1.4237e-04	 & 4.7131e-05 \\
\cacador	 & \jager	 & 1.0000e+00	 & 1.0000e+00	 & 0.0000e+00 \\
\cacador	 & \searcher	 & 9.5401e-23	 & 8.9041e-22	 & 1.3045e-03 \\
\cyobstract	 & \iocextract	 & 1.0013e-02	 & 1.2744e-02	 & 2.4565e-04 \\
\cyobstract	 & \iocextractor	 & 1.1071e-01	 & 1.2916e-01	 & 1.0536e-04 \\
\cyobstract	 & \iocfinder	 & 1.1717e-01	 & 1.3123e-01	 & 9.9363e-05 \\
\cyobstract	 & \iocparser	 & 4.2949e-04	 & 7.5162e-04	 & 6.7494e-04 \\
\cyobstract	 & \jager	 & 2.1251e-02	 & 2.5871e-02	 & 3.6767e-04 \\
\cyobstract	 & \searcher	 & 5.4571e-03	 & 8.0421e-03	 & 2.8976e-04 \\
\iocextract	 & \iocextractor	 & 5.9633e-03	 & 8.3486e-03	 & 2.9454e-05 \\
\iocextract	 & \iocfinder	 & 3.5693e-03	 & 5.5523e-03	 & 3.2923e-05 \\
\iocextract	 & \iocparser	 & 3.6201e-25	 & 5.0681e-24	 & 1.7284e-03 \\
\iocextract	 & \jager	 & 2.6479e-10	 & 7.4141e-10	 & 1.2132e-03 \\
\iocextract	 & \searcher	 & 2.4354e-08	 & 5.2456e-08	 & 1.8189e-06 \\
\iocextractor	 & \iocfinder	 & 6.2505e-01	 & 6.4820e-01	 & 9.3784e-08 \\
\iocextractor	 & \iocparser	 & 3.8698e-22	 & 2.7089e-21	 & 1.3122e-03 \\
\iocextractor	 & \jager	 & 1.1041e-08	 & 2.8105e-08	 & 8.6746e-04 \\
\iocextractor	 & \searcher	 & 6.1777e-04	 & 1.0175e-03	 & 4.5934e-05 \\
\iocfinder	 & \iocparser	 & 1.4453e-21	 & 8.0938e-21	 & 1.2927e-03 \\
\iocfinder	 & \jager	 & 1.3778e-08	 & 3.2149e-08	 & 8.5116e-04 \\
\iocfinder	 & \searcher	 & 3.6823e-04	 & 6.8737e-04	 & 5.0259e-05 \\
\iocparser	 & \jager	 & 1.3860e-01	 & 1.4927e-01	 & 4.7302e-05 \\
\iocparser	 & \searcher	 & 1.0825e-26	 & 3.0310e-25	 & 1.8419e-03 \\
\jager	 & \searcher	 & 5.4592e-11	 & 1.6984e-10	 & 1.3090e-03 \\
  \hline
  \end{tabular}
\end{table}

\paragraph{Statistical significance.}
We evaluate whether differences in the extracted indicators across tools are statistically significant by first 
performing an analysis of variance test and 
then measuring the strength of the differences
using pairwise t-tests between the tools.
The independent variable is the tool used to produce each set of IOC detections.
There are eight groups accounting for the source of variability, one per tool.
The dependent variable is the number of detections (TPs) produced by each tool.
We perform a Repeated Measures One-Way ANOVA test 
because all tools are evaluated on the same samples (\ie documents).
We focus on the 6 indicator types extracted by all tools
(\ioc{email}, \ioc{ip4}, \ioc{md5}, \ioc{sha1}, \ioc{sha256}, \ioc{url}).

\begin{figure}[t]
  \centering
      \includegraphics[width=\columnwidth]{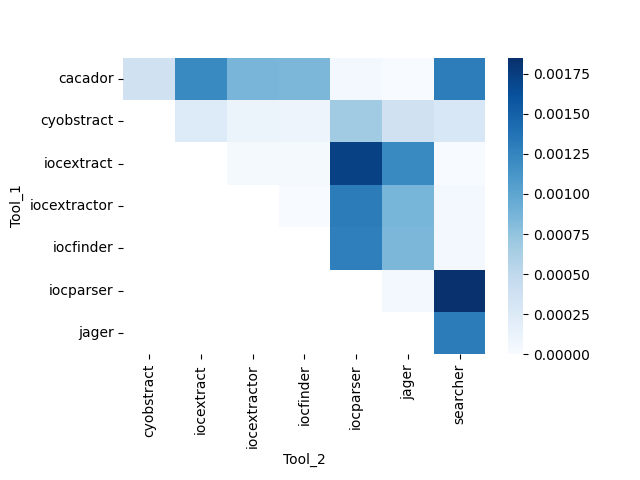}
  \caption{Eta-square values resulting from pairwise comparison of tools.}
  \label{fig:anova}
\end{figure}

The test results indicate a significant statistical difference
between the detections produced by the tools,
with a p-value lower than $10^{-5}$.
We apply pairwise t-tests among groups
to identify which pairs of tools exhibit significant differences.
We use the Benjamini/Hochberg FDR correction method for the p-values
because of the multiple statistical tests produced.
Table~\ref{tbl:anova} shows the results of this test.
A statistically significant difference between detections
is found among every pair of tools (shown by the \textit{p-corr} column), 
except for 5 pairs 
(\cacador--\jager,
\cyobstract--\iocextractor, 
\cyobstract--\iocfinder,
\iocextractor--\iocfinder, and 
\iocparser--\jager).
Figure~\ref{fig:anova} shows the strength of the differences between tools,
measured by the \textit{eta-square} value resulting from the tests.
The results show that detections produced by \searcher
are statistically significant compared to all other tools.

\begin{table}[t]
\scriptsize
\centering
\caption{Indicator extraction runtime (in seconds) across all APTnotes and 
average by indicator type.}
\label{tbl:runtime}
\begin{tabular}{lrrr}
\hline
\textbf{Tool} & \textbf{Runtime (sec)} & \textbf{Ind.} & \textbf{Avg. per Ind. (sec)} \\
\hline
	\cacador		&   489.23 & 12 &   40.76 \\
	\iocextract	&  1350.69 &  8 &  168.83 \\
	\iocparser	&  1563.33 & 11 &  142.12 \\
	\iocextractor	&  1978.51 & 18 &  109.91 \\
	\jager		&  2360.02 & 11 &  214.54 \\
	\searcher	&  2847.98 & 41 &   69.46 \\
	\cyobstract	&  6945.03 & 25 &  277.80 \\
	\iocfinder	& 96596.23 & 25 & 3863.84 \\

	\ignore{
cacador		& 253.2	& 12	& 21.1 \\
ioc\_parser	& 439.3	& 11	& 39.9 \\
ioc-extractor	& 586.6	& 18	& 32.6 \\
jager		& 642.0	& 11	& 58.4 \\
iocextract	& 699.4	&  8	& 87.4 \\
searcher	& 831.9	& 41	& 20.3 \\
cyobstract	& 1,960.3	& 25	& 78.4 \\
ioc-finder	& 53,807.3	& 25	& 2,152.3\\
	}
\hline
\end{tabular}
\end{table}

\paragraph{Runtime.}
Table~\ref{tbl:runtime} shows the total time in seconds each tool took to extract indicators on the \numcomparisonreports text documents, 
the number of indicator types supported, and the average time by indicator type.
In our tests, the fastest tool is \cacador, likely due to being implemented in Go and compiled to native code.
The slowest tool is \iocfinder, which is 14 times slower than the second slowest tool (\cyobstract).
This is probably because \iocfinder is the only tool that uses grammars for the
extraction, highlighting the efficiency of using \regexps to extract indicators.
All tools perform one pass on the text for each \regexp and they typically use
one \regexp for each indicator type.
Thus, the runtime is highly influenced by the number of indicator types
extracted.  When normalizing the total runtime by the number of extracted
indicator types, \cacador is still the fastest with 40.8 seconds, followed by
\searcher (69.5), and \iocextractor (109.9).

\begin{table}[t]
\scriptsize
\centering
\caption{Social indicators validation.}
\label{tbl:social}
\begin{tabular}{lrr}
\hline
\textbf{Indicator} & \textbf{Total} & \textbf{Validated} \\
\hline
 {facebookHandle} & 150 & 124 \\
   {gitHubHandle} & 258 & 242 \\
{instagramHandle} &  10 &   8 \\
{pinterestHandle} &   1 &   1 \\
  {youTubeHandle} &  24 &  24 \\
 {youTubeChannel} &   6 &   6 \\
  {twitterHandle} & 516 & 431 \\
\hline
All & 965 & 836 \\
\hline
\end{tabular}
\end{table}

\paragraph{Social handle validation.}
Our majority-based evaluation cannot be applied to the 38 indicators 
only extracted by one tool.
Among those, there are 11 social indicators only 
extracted by \searcher, of which 9 appear in \aptnotes and \chainsmith reports
(no Telegram and WhatsApp handles are found in these sources).
For social network handles we perform an alternative validation, 
which checks whether there \textit{currently} exists an account in the social
network for that handle.
For this, we use the Blackbird~\cite{blackbird_github} open-source tool,
which given a username checks if there currently exists an account with 
that handle on 143 different social networks.
Blackbird supports 6 social networks \searcher extracts handles for
(Facebook, GitHub, Instagram, Pinterest, Telegram, and YouTube). 
In addition, we expanded Blackbird to also support 
\ioc{youtubeChannel} indicators.
For \twitter handles, we leverage the official Twitter API 
for the same purpose~\cite{twitter_official_api}.
LinkedIn does not allow searching for account names. 

Table~\ref{tbl:social} summarizes the results.
Our automated approach successfully validated 87\% of the social indicators.
We manually inspect the 102 indicators that we could not validate with Blackbird, 
using a web browser to search for a particular handle on each social network.
For Twitter, 15 of of the 85 usernames that did not validate, were linked to suspended accounts (e.g., \url{@MalwareSigs}).
In all of the remaining cases, the social networks returned a page to inform us that the particular account does not exist.
For 75 of the 87 accounts for which we could not find the user profile page, 
the identifier contained a meaningful sequence of characters 
(e.g., ``sucuri\_security'', ``avast\_antivirus''), 
suggesting that the identifier could belong to an old account that was closed.
Overall, this evaluation suggests 
that over 90\% of the social identifiers \searcher extracted are 
likely true positives. For the rest, we are not able to determine 
if they are false positives or correspond to accounts that 
have been closed since they appeared in the sources.

\begin{table}[t]
\scriptsize
\centering
\caption{Accuracy of filtering approaches on a manually-built ground truth.}
\label{tbl:filterEval}
\begin{tabular}{l|r|r|r|r|r|r|r}
\hline
	\textbf{Tool} & \textbf{TP} & \textbf{FP} & \textbf{FN} & \textbf{TN} & \textbf{Prec.} & \textbf{Recall} & \textbf{F1} \\
\hline
\cacador		 &  29 & 69 & 0 & 8 & 0.30 & 1.00 & 0.46 \\
\iocparser	 &  29 & 66 & 0 & 11 & 0.31 & 1.00 & 0.47 \\
\searcher	   &  29 & 6 & 0 & 71 & 0.83 & 1.00 & 0.91 \\
\hline
\end{tabular}
\end{table}

\paragraph{Filtering.}
To evaluate the filtering, we manually built a ground truth (GT) 
of \numgt indicators.
We randomly selected those indicators among those extracted by 
\searcher across all datasets and prior to the filtering.
Then, an analyst manually reviewed the documents from where the indicators 
were extracted to determine whether each indicator was benign or malicious.
The analyst made a determination based on the indicator context 
in the document, i.e., by reading the parts of the document that
mentioned the indicator.

To measure the filtering accuracy, we first apply the filtering approach of
\cacador, \iocparser, and our platform to the \numgt indicators.  Then, for each
approach, we compare the indicators that survive the filtering with the GT.  We
consider a true positive (TP) a malicious indicator in the GT that correctly
survives the filtering and a TN is a benign indicator in the GT that was
correctly filtered out.  A false positive (FP) is a benign indicator in the GT
that incorrectly passes the filtering, while a false negative (FN) is a
malicious indicator in the GT incorrectly removed by the filtering.
From those counts, we compute the precision, recall, and F1 score. 
Note that it does not matter that indicators in the GT were selected among 
those extracted by \searcher, as we do not extract indicators using 
\cacador and \iocparser in this experiment. 
We only run their filtering rules on the GT indicators. 

Table~\ref{tbl:filterEval} summarizes the results.
All three filtering approaches have zero FNs and thus perfect recall, 
i.e., they do not remove any truly malicious indicators.
On the other hand, precision significantly differs;
our filtering achieves more than twice higher precision (0.83)
than the filtering used by \iocparser (0.31) and \cacador (0.30).
The low precision by \cacador and \iocparser is due to its filtering being 
fairly incomplete,
allowing many benign indicators to be output as malicious IOCs. 
The final F1 score is 0.91 for our filtering, 
followed by \iocparser (0.47) and \cacador (0.46).
The results indicate that the static blocklists used by \iocparser and 
\cacador fail to filter many of the benign indicators in the documents. 
Furthermore, despite the \iocparser blocklist being an order of magnitude 
larger than the \cacador blocklist, 
the recall improvement is marginal.
This shows that it is hard to predict the sources from where a user will 
collect reports, making static blocklists largely incomplete.
The use of a dynamic blocklist in our filtering module  
significantly improves the recall by adjusting the filtering to the 
collected sources.
For example, it is very common for tweets to contain URLs to external references,
including links to documents from other sources (e.g, an already monitored blog or RSS feed).
The blocklists of \cacador and \iocparser would assume that such a URL is an IOC, possibly because the tool developers did not consider Twitter as a source.
\Searcher correctly flags those generic indicators as FPs, since it relies on a blocklist that is dynamically expanded each time the user adds a new source that will be monitored.
In summary, the dynamic blocklist allows filtering a larger amount of benign indicators than static blocklists without losing any malicious IOCs.

\section{Discussion}
\label{sec:discussion}

This section discusses limitations and future improvements.

\paragraph{Threats to validity.}
Our methodology for comparing indicator extraction tools assumes that the 
majority of tools will be correct. 
However, in some cases, a minority of tools may be correct instead, 
as illustrated by the \ioc{bitcoin} results 
where \searcher is wrongly assigned FNs due to FPs from two other tools. 
In general, the more tools support an indicator, 
the higher confidence we have in the majority of results.
An alternative approach for evaluation tools is to use a ground truth (GT) 
dataset.
However, the manual effort required to build a GT with thousands of documents 
would be very large, and 
it would be difficult to cover different indicator types.
For this reason, \regexp evaluations tend to be built leveraging 
synthetic examples (e.g.,~\cite{urlregexp}). 
Unfortunately, synthetic examples may not represent results on real documents.

Our evaluation could be biased due to document selection, 
unknown normalizations, expired social accounts, and 
the size of the ground truth used for evaluating the filtering.
For example, we evaluate tools exclusively on English documents.
While \regexps should largely be language-agnostic, it is possible that 
some language-specific feature (e.g., special characters) affects the results.
It is also possible for our comparison methodology to negatively affect 
a tool if we miss some normalization it performs on its outputs.
The validation of social handles may underestimate the \searcher accuracy since
extracted handles may correspond to accounts since deleted.
And, our filtering evaluation is performed on a small ground truth of 
106 indicators. 
A larger ground truth may uncover harder cases where accuracy drops. 

\paragraph{Adding origins.}
Adding new \origins is currently the only step where \tool 
requires human involvement.
\tool can automatically identify new origins
to be considered for inclusion.
For this, it examines the report URLs in the crowd-sourced datasets 
(\malpedia, \aptnotes) 
ranking domains in the report URLs by the number of report URLs 
where they appear. 
Then, it filters domains already in the list of RSS origins. 
Finally, the top-ranked domains are flagged as potential cybersecurity 
blogs, so that an analyst can examine them and 
identify their RSS URL.
To identify candidate new Twitter accounts to monitor, 
\tool examines the re-tweets in the monitored accounts, building a ranking 
from the original tweeting account to the number of collected re-tweets 
from the account.
The top re-tweeted accounts, not yet monitored, are output 
so that a human can analyze them for inclusion. 
In future work, we would like to explore integrating techniques to 
identify valuable Twitter accounts by analyzing their posts~\cite{iocminer}.  

\paragraph{Report variants.}
\updated{
If the same document (i.e., same SHA256 hash) is collected from
different origins, \tool deduplicates it to store only one copy of the 
document and updates the document's traceability information 
to capture all origins from where the document was collected. 
As introduced in Section~\ref{sec:collection}, 
it is possible that \tool collects multiple documents 
corresponding to different instances of the same threat report.
This may happen if a report is distributed in different formats, 
in case \tool downloads multiple times the URL of a report that points to a 
webpage with dynamic content, 
and if a report generates multiple versions over time 
(e.g., by fixing some errata).
Currently, the user can identify multiple instances of the same report by 
querying for all documents downloaded from the same URL or 
for all documents with the same title.
However, this may miss some cases, e.g., when the title has changed.
To address this issue, \tool could store a similarity hash of the 
text content (extracted from the PDF or HTML document), 
which could be used to identify small variations of the same content. 

In this work, \tool was configured to download a URL only once
(unless an error happened).
This configuration reduces the number of needed downloads and 
removes some of the above cases. 
On the other hand, it may fail to collect different versions of the same report 
that appear over time.
If the user wants to collect multiple versions of a report, 
he can easily change this configuration, 
at the cost of increased network bandwidth and runtime.
In future work, we would like to explore leveraging the 
ETag header to keep track of the document pointed by a URL and 
use an HTTP HEAD request to check if the document was updated
since the last time we retrieved it.
With this approach, the report 
would only be re-downloaded if a new version exists. 
However, not all websites provide the ETag header or support 
HTTP HEAD requests.
}

\section{Conclusions}
\label{sec:conclusions}

This paper presents \tool, 
an automated platform for collecting threat reports from 6 sources 
(\rss, \twitter, \telegram, \malpedia, \aptnotes, \chainsmith) and 
comparing the accuracy of indicator extraction tools on the collected reports.
\Tool implements a novel majority vote methodology 
for comparing the accuracy of indicator extraction tools,
which does not require a manually-built ground truth.
\Tool continuously monitors the sources, 
downloads new threat reports,
extracts indicators from the reports, and 
filters generic indicators to produce a list of IOCs.
\Tool includes the \searcher tool for extracting \numiocs indicator types 
from HTML, PDF, and text files using \regexps.
We run \tool for over 15 months to collect \numentries reports from 
the 6 sources; 
extract \numextractedindicators indicators from the reports using \searcher; 
and identify \numextractediocs IOCs.
We analyze the collected data to identify the top IOC contributors and the IOC class distribution.
Then, we applied \tool for comparing 7 popular indicator extraction tools 
with \searcher and assess their accuracy using our novel 
majority-vote methodology.

\section*{Acknowledgments}
We are grateful to Pierre Ganty for his help with the rat 
ReDoS checker tool.
This work has been partially supported by the Madrid regional government 
through program S2018/TCS-4339 (BLOQUES-CM) and 
Atracción de Talento grant 2020-T2/TIC-20184.
This work has also been partially supported by 
MCIN/AEI/10.13039/501100011033/ through grants 
RTI2018-102043-B-I00 (SCUM),
TED2021-132464B-I00 (PRODIGY), and 
PRE2019-088472. 
Partial funding was also provided by 
Ministerio de Ciencia, Innovación y Universidades grant FPU18/06416.
The above programs are co-funded by 
the European Union EIE, ESF, and NextGeneration EU/PRTR Funds.
Any opinions, findings, and conclusions or recommendations expressed in 
this material are those of the authors or originators, and 
do not necessarily reflect the views of the sponsors.

\bibliographystyle{elsarticle-num}
\bibliography{paper}

\begin{thebibliography}{10}
\expandafter\ifx\csname url\endcsname\relax
  \def\url#1{\texttt{#1}}\fi
\expandafter\ifx\csname urlprefix\endcsname\relax\def\urlprefix{URL }\fi
\expandafter\ifx\csname href\endcsname\relax
  \def\href#1#2{#2} \def\path#1{#1}\fi

\bibitem{timarket}
MarketWatch, Threat intelligence market 2022 report,
  \url{https://www.marketwatch.com/press-release/threat-intelligence-market-2022-report-examines-latest-trends-and-key-drivers-supporting-growth-till-2030-2022-07-27}
  (2022).

\bibitem{tealeaves}
V.~G. Li, M.~Dunn, P.~Pearce, D.~McCoy, G.~M. Voelker, S.~Savage, {Reading the
  Tea leaves: A Comparative Analysis of Threat Intelligence}, in: USENIX
  Security, 2019.

\bibitem{bouwman2020different}
X.~Bouwman, H.~Griffioen, J.~Egbers, C.~Doerr, B.~Klievink, M.~Van~Eeten, {A
  different cup of TI? The added value of commercial threat intelligence}, in:
  USENIX Security Symposium, 2020.

\bibitem{stix}
O.~Open, Stix: A structured language for cyber threat intelligence,
  \url{https://oasis-open.github.io/cti-documentation/} (2022).

\bibitem{openioc}
W.~Gibb, D.~Kerr, Openioc: Back to the basics,
  \url{https://www.mandiant.com/resources/openioc-basics} (October 2013).

\bibitem{sabottke_usenix2015}
C.~Sabottke, O.~Suciu, T.~Dumitras, {Vulnerability Disclosure in the Age of
  Social Media: Exploiting Twitter for Predicting Real-World Exploits}, in:
  USENIX Security, 2015.

\bibitem{alves_esorics2020}
F.~Alves, A.~Andongabo, I.~Gashi, P.~M. Ferreira, A.~Bessani, {Follow the Blue
  Bird: {A} Study on Threat Data Published on Twitter}, in: ESORICS, 2020.

\bibitem{ssdeep}
J.~Kornblum, {ssdeep Project},
  \url{https://ssdeep-project.github.io/ssdeep/index.html} (2016).

\bibitem{jager}
S.~J. Roberts, {jager}, \url{https://github.com/sroberts/jager} (2015).

\bibitem{iocparser}
A.~Buescher, {ioc\_parser}, \url{https://github.com/armbues/ioc_parser/}
  (2017).

\bibitem{cacador}
S.~J. Roberts, {Cacador}, \url{https://github.com/sroberts/cacador} (2016).

\bibitem{cyobstract}
M.~Sisk, R.~Ruefle, S.~Perl, {Harvesting Artifacts: Improving Useful Data
  Extraction from Cybersecurity Incident Reports},
  \url{https://github.com/cmu-sei/cyobstract} (2018).

\bibitem{iocextract}
InQuest, {iocextract}, \url{https://github.com/InQuest/python-iocextract}
  (2019).

\bibitem{iocextractor}
M.~Niseki, {IoC extractor}, \url{https://github.com/ninoseki/ioc-extractor}
  (2019).

\bibitem{liao2016acing}
X.~Liao, K.~Yuan, X.~Wang, Z.~Li, L.~Xing, R.~Beyah, {Acing the IOC Game:
  Toward Automatic Discovery and Analysis of Open-Source Cyber Threat
  Intelligence}, in: CCS, 2016.

\bibitem{ttpdrill}
G.~Husari, E.~Al-Shaer, M.~Ahmed, B.~Chu, X.~Niu, {TTPDrill: Automatic and
  Accurate Extraction of Threat Actionsfrom Unstructured Text of CTI Sources},
  in: ACSAC, 2017.

\bibitem{chainsmith}
Z.~Zhu, T.~Dumitras, {ChainSmith: Automatically Learning the Semantics of
  Malicious Campaigns by Mining Threat Intelligence Reports}, in: Euro S{\&}P,
  2018.

\bibitem{satvat2021extractor}
K.~Satvat, R.~Gjomemo, V.~Venkatakrishnan, {Extractor: Extracting Attack
  Behavior from Threat Reports}, in: Euro S{\&}P, 2021.

\bibitem{davis2018impact}
J.~C. Davis, C.~A. Coghlan, F.~Servant, D.~Lee, {The Impact of Regular
  Expression Denial of Service (ReDoS) in Practice: An Empirical Study at the
  Ecosystem Scale}, in: ACM Joint Meeting on European Software Engineering
  Conference and Symposium on the Foundations of Software Engineering, 2018.

\bibitem{malpedia}
D.~Plohmann, M.~Clauß, S.~Enders, E.~Padilla, {Malpedia: A Collaborative
  Effort to Inventorize the Malware Landscape}, The Journal on Cybercrime \&
  Digital Investigations 3~(1) (2017).

\bibitem{aptnotes}
K.~Bandla, S.~Castro, {APTnotes}, \url{https://github.com/aptnotes/data}
  (2022).

\bibitem{iocfinder}
F.~Hightower, {IOC Finder}, \url{https://github.com/fhightower/ioc-finder}
  (2018).

\bibitem{iocminer}
A.~Niakanlahiji, L.~Safarnejad, R.~Harper, B.-T. Chu, {IoCMiner: Automatic
  Extraction of Indicators of Compromise from Twitter}, in: IEEE Big Data,
  2019.

\bibitem{timiner}
J.~Zhao, Q.~Yan, J.~Li, M.~Shao, Z.~He, B.~Li, {TIMiner: Automatically
  extracting and analyzing categorized cyber threat intelligence from social
  data}, Computers \& Security (2020).

\bibitem{twiti}
H.~Shin, W.~Shim, S.~Kim, S.~Lee, Y.~G. Kang, Y.~H. Hwang, {Twiti: Social
  Listening for Threat Intelligence}, in: WWW, 2021.

\bibitem{caballero_jnic2022}
J.~Caballero, G.~Gomez, S.~Matic, G.~Sánchez, S.~Sebastián, A.~Villacañas,
  {(Work-in-progress) FATR: a Framework for Automated Analysis of Threat
  Reports}, in: JNIC, 2022.

\bibitem{selenium}
{Software Freedom Conservancy}, Selenium, \url{https://www.selenium.dev/}
  (2022).

\bibitem{python_requests}
{Python Software Foundation}, {Requests}, \url{https://github.com/psf/requests}
  (2022).

\bibitem{iocsearcher}
MaliciaLab, {iocsearcher}, \url{https://github.com/malicialab/iocsearcher}
  (2019).

\bibitem{privee}
S.~Zimmeck, S.~M. Bellovin, {Privee: An Architecture for Automatically
  Analyzing Web Privacy Policies}, in: USENIX Security, 2014.

\bibitem{slavin2016toward}
R.~Slavin, X.~Wang, M.~B. Hosseini, J.~Hester, R.~Krishnan, J.~Bhatia, T.~D.
  Breaux, J.~Niu, {Toward a Framework for Detecting Privacy Policy Violations
  in Android Application Code}, in: International Conference on Software
  Engineering, 2016.

\bibitem{zimmeck2017automated}
S.~Zimmeck, Z.~Wang, L.~Zou, R.~Iyengar, B.~Liu, F.~Schaub, S.~Wilson, N.~M.~S.
  M, S.~M. Bellovin, J.~R. Reidenberg, {Automated Analysis of Privacy
  Requirements for Mobile Apps}, in: NDSS, 2017.

\bibitem{polisis}
H.~Harkous, K.~Fawaz, R.~Lebret, F.~Schaub, K.~G. Shin, K.~Aberer, {Polisis:
  Automated Analysis and Presentation of Privacy Policies Using Deep Learning},
  in: USENIX Security, 2018.

\bibitem{policylint}
B.~Andow, S.~Y. Mahmud, W.~Wang, J.~Whitaker, W.~Enck, B.~Reaves, K.~Singh,
  T.~Xie, {PolicyLint: Investigating Internal Privacy Policy Contradictions on
  Google Play}, in: USENIX Security, 2019.

\bibitem{pdfminer}
Y.~Shinyama, P.~Guglielmetti, P.~Marsman, {pdminer.six},
  \url{https://github.com/pdfminer/pdfminer.six} (2019).

\bibitem{bs4}
L.~Richardson, {Beautiful Soup},
  \url{https://beautiful-soup-4.readthedocs.io/en/latest/} (2022).

\bibitem{readability}
Mozilla, {Mozilla Readability.js library},
  \url{https://github.com/mozilla/readability} (2022).

\bibitem{hosseini21unifying}
H.~Hosseini, M.~Degeling, C.~Utz, T.~Hupperich, {Unifying Privacy Policy
  Detection}, in: PoPETs, 2021.

\bibitem{boilerpipe}
C.~Kohlschütter, {boilerpipe: Boilerplate Removal and Fulltext Extraction from
  HTML pages}, \url{https://github.com/kohlschutter/boilerpipe} (2022).

\bibitem{gao2021enabling}
P.~Gao, F.~Shao, X.~Liu, X.~Xiao, Z.~Qin, F.~Xu, P.~Mittal, S.~R. Kulkarni,
  D.~Song, {Enabling Efficient Cyber Threat Hunting With Cyber Threat
  Intelligence}, in: IEEE International Conference on Data Engineering, 2021.

\bibitem{webmoney}
{WebMoney}, {WebMoney - Universal Payment System},
  \url{https://www.wmtransfer.com/} (2022).

\bibitem{tlp}
{FIRST}, {Traffic Light Protocol (TLP). FIRST Standards Definitions and Usage
  Guidance - Version 1.0}, \url{https://www.first.org/tlp/docs/tlp-v1.pdf}
  (2016).

\bibitem{imphash}
Mandiant, {Tracking Malware with Import Hashing},
  \url{https://www.mandiant.com/resources/tracking-malware-import-hashing}
  (2014).

\bibitem{trancondss2019}
V.~L. Pochat, T.~V. Goethem, S.~Tajalizadehkhoob, M.~Korczy\'{n}ski, W.~Joosen,
  {Tranco: A Research-Oriented Top Sites Ranking Hardened Against
  Manipulation}, in: NDSS, 2019.

\bibitem{btcrelations}
G.~Gomez, P.~Moreno-Sanchez, J.~Caballero, {Watch Your Back: Identifying
  Cybercrime Financial Relationships in Bitcoin through Back-and-Forth
  Exploration}, in: CCS, 2022.

\bibitem{iocextractRedos}
{DaveCrim}, {Catastrophic backtracking in BACKSLASH\_URL\_RE},
  \url{https://github.com/InQuest/python-iocextract/issues/52} (2021).

\bibitem{ratRedos}
F.~Parolini, A.~Min{\'e}, rat - redos abstract tester,
  \url{https://github.com/parof/rat} (April 2022).

\bibitem{parolini2022sound}
F.~Parolini, A.~Min{\'e}, {Sound Static Analysis of Regular Expressions
  for Vulnerabilities to Denial of Service Attacks}, in: International
  Symposium on Theoretical Aspects of Software Engineering, 2022.

\bibitem{blackbird_github}
{p1ngul1n0}, {Blackbird}, \url{https://github.com/p1ngul1n0/blackbird} (2022).

\bibitem{twitter_official_api}
{Twitter}, {Twitter API},
  \url{https://developer.twitter.com/en/docs/twitter-api} (2022).

\bibitem{urlregexp}
{Mathias Bynens}, {In search of the perfect URL validation regex},
  \url{https://mathiasbynens.be/demo/url-regex} (2021).

\end{thebibliography}

\end{document}